\documentclass[twocolumn]{autart}    

\usepackage{graphicx}          
\usepackage{amsmath} 
\usepackage{amssymb}  
\usepackage{url}
\usepackage{algorithm}
\usepackage{algpseudocode}
\usepackage{xcolor}
\usepackage{accents}

\def\awinr{\mc{R}_A} 
\def\awinri{\mc{R}_I} 
\def\awinrc{\mc{R}_C} 
\def\dwinr{\mc{R}_D} 
\def\dwinrp{\mc R_\text{pair}}

\def\lp{\left(}
\def\rp{\right)}

\def\target{\mc{T}}

\def\numcapture{N_A^\text{cap}}

\def\acontrol{\mf{u}_A}

\def\Qmm{Q_\text{MM}}
\def\Qmis{Q_\text{MIS}}

\def\controld{\omega}

\def\curve{\bs{\gamma}}
\def\tangent{\bs{T}}
\def\acos{\cos^{-1}}

\def\sopt{s_\text{opt}}
\def\smid{s_\text{mid}}

\def\xa{\mf x_A}
\def\di{D_i}
\def\dj{D_j}
\def\sdi{s_{\di}}
\def\sdj{s_{\dj}}
\def\stanl{s_\text{tan,L}}
\def\stanr{s_\text{tan,R}}
\def\objective{P}
\def\objectivea{\objective_{1}}
\def\objectiveb{\objective_{2}}
\def\sDoppo{s_D^\text{op}}
\def\sdirect{\mc{S}_\text{d}}
\def\sleft{\mc{S}_\text{L}}
\def\sright{\mc{S}_\text{R}}
\def\singsurface{\bs{\Gamma}}
\def\affsurface{\bs{\Gamma}_\text{aff}}
\def\dispsurface{\bs{\Gamma}_\text{dis}}

\def\Rleft{\mc{R}_L}

\def\distbarrier{d_\text{bar}}
\def\rotmat{\mf{R}}

\newcommand{\unitvec}[2]{\hat{\mf{x}}_{#1\veryshortarrow#2}}
\newcommand{\relvec}[2]{\mf{x}_{#1\veryshortarrow#2}}
\newcommand{\arclength}[2]{s_{#1\veryshortarrow#2}}
\newcommand{\segment}[2]{[#1,#2]}
\newcommand{\veryshortarrow}[1][3pt]{\mathrel{%
   \hbox{\rule[\dimexpr\fontdimen22\textfont2-.2pt\relax]{#1}{.4pt}}%
   \mkern-4mu\hbox{\usefont{U}{lasy}{m}{n}\symbol{41}}}}

\def\bs{\boldsymbol}
\def\mf{\mathbf}

\def\mc{\mathcal}
\def\beq{\begin{equation*}}
\def\eeq{\end{equation*}}
\def\bql{\begin{equation}}
\def\eql{\end{equation}}
\def\bqn{\begin{eqnarray*}}
\def\eqn{\end{eqnarray*}}
\def\bnl{\begin{eqnarray}}
\def\enl{\end{eqnarray}}

\newcommand{\rev}[1]{\textcolor{black}{#1}}
\newcommand{\revrev}[1]{\textcolor{black}{#1}}

\DeclareMathOperator*{\argmax}{arg\,max}
\DeclareMathOperator*{\argmin}{arg\,min}

\newtheorem{theorem}{Theorem}
\newtheorem{lemma}{Lemma}
\newtheorem{definition}{Definition}
\newtheorem{remark}{Remark}

\def\ARXIVversion{1}

\def\maincolor{black}
\def\arxivcolor{black}

\begin{document}

\begin{frontmatter}

\title{Perimeter-defense~Game on Arbitrary Convex Shapes} 

\thanks[footnoteinfo]{
We gratefully acknowledge the support of ARL grant ARL DCIST CRA W911NF-17-2-0181.
Corresponding author: D.~Shishika. 
}

\author[GMU]{Daigo Shishika}\ead{dshishik@gmu.edu},    
\author[Penn]{Vijay Kumar}\ead{kumar@seas.upenn.edu}  

\address[GMU]{Department of Mechanical Engineering, George Mason University, USA.}  
\address[Penn]{GRASP Lab, University of Pennsylvania, USA.}  


\begin{abstract}                          
\if \ARXIVversion1
This paper studies a variant of multi-player reach-avoid game played between intruders and defenders. The intruder team tries to score by sending as many intruders as possible to the target area, while the defender team tries to minimize this score by intercepting them. Specifically, we consider the case where the defenders are constrained to move on the perimeter of the target area. Since it is challenging to directly solve the multi-player game due to the high dimensionality of the joint state space, we leverage the solutions to smaller scale problems. First, we solve the one vs.\ one game, for which existing works either rely on numerical approaches or make simplifying assumptions (e.g., circular perimeter, or equal speed). This paper accommodates target areas with any arbitrary convex shapes and provides analytical solution which lends itself to a useful geometric interpretation. We also provide a detailed discussion on the optimality of the derived strategies. Secondly, we solve the two vs.\ one game to introduce a cooperative pincer maneuver, where a pair of defenders team up to capture an intruder that cannot be captured by either one of the defender individually. Finally, we introduce how the aforementioned building blocks are used in three different assignment-based defense strategies.
\fi
\if \ARXIVversion0
This paper studies a variant of multi-player reach-avoid game played between intruders and defenders. The intruder team tries to score by sending as many intruders as possible to the target area, while the defender team tries to minimize this score by intercepting them. Specifically, we consider the case where the defenders are constrained to move on the perimeter of the target area. 
In this work, we first solve the one vs.\ one game, for which existing works either rely on numerical approaches or make simplifying assumptions (e.g., circular perimeter, or equal speed). This paper accommodates target areas with any arbitrary convex shapes and provides analytical solution which lends itself to a useful geometric interpretation. Secondly, we solve the two vs.\ one game to introduce a cooperative pincer maneuver, where a pair of defenders team up to capture an intruder that cannot be captured by either one of the defender individually. 
We also provide a detailed discussion on the optimality of the derived strategies.
The results of this paper directly feed into different assignment-based defense strategies that exist in the literature.
\fi
\end{abstract}

\end{frontmatter}

\section{Introduction}
Maintaining perimeter surveillance and security is a complex problem given that it has become practical to deploy autonomous agents in large numbers.
Various approaches to counter intrusions by unmanned vehicles have been studied including patrolling strategy~\cite{Pasqualetti2012}, intrusion detection based on behavior rules~\cite{Mitchell2014}, and GPS spoofing to manipulate the behavior of the agents~\cite{Kerns2014}.

When evasive targets need to be detected, intercepted, or surrounded, the scenarios are often formulated as pursuit-evasion games
(PEGs)~\cite{Chung2011,garcia2020multiple,von2019multi}.
If it is formulated as a \emph{game of kind}, we ask which initial configuration leads to capture (or evasion), and what pursuit (or evasive) strategy guarantees that.
If it is formulated as a \emph{game of degree}, we find the optimal strategy for a given objective function, e.g., time to capture.

The game becomes more complex when the evader has another objective, such as to reach a target.
A version of this problem is called the target-attacker-defender (TAD) game \cite{Garcia2015,Liang2019,Rubinsky2014}.
In a TAD game the attacker aims to capture the target while avoiding being captured by the defender, and the defender tries to defend the target by intercepting the attacker.
In \cite{Liang2019} the defender could save the target by reaching it before the attacker, which led to a rendezvous type strategy.

Another formulation focuses on the case where the target is a region in the game space and is no longer treated as an agent.
The two-player version of the game (one defender vs. one attacker) was first introduced by Isaacs as the target-defense game \cite{Isaacs}.
\rev{This game is also called the reach-avoid game \cite{Fisac2015a,zhou2018efficient,Zhou2012general}, and it has been studied in many different variants including multi-player scenario \cite{Chen2014a,Huang2011icra,yan2017escape,yan2018twovsone,yan2019analytical} and coast-line guarding or boarder defense \cite{garcia2019cooperative,garcia2019coastline,von2019multiple}}.

This paper considers the perimeter defense game, which is a variant of the reach-avoid game played between intruders and defenders \cite{shishika2018cdc,shishika2020ral}.
The intruder team tries to score by sending as many intruders as possible to the target area, while the defender team tries to minimize this score by intercepting them.
A specific assumption made in this paper is that the defenders are constrained to move on the perimeter.
Such assumption is motivated by the scenarios where the target region acts as an obstacle that the defenders cannot move through
( e.g., defending a perimeter of a building using ground vehicles).

Various solution methods have been proposed to solve the PEGs introduced thus far.
In general the approaches can be divided into two types: the differential game formulation and the explicit policy method \cite{Liang2019}.
The former obtains the strategies and the winning regions by solving a Hamilton-Jacobi-Issacs (HJI) partial differential equation (PDE), while the latter analyzes the outcome of the game by prescribing a strategy to the players.

The differential game formulation has been successfully utilized for relatively simple problems that allow analytical solution to the HJI PDEs \cite{TamerBasar,Isaacs,Makkapati2018} and other problems with low dimensional state space for which the HJI PDEs can be numerically solved \cite{Chen2014a,Fisac2015}.
The strength of this approach is that the optimality of the derived strategies are ensured by construction.
The down side is the curse of dimensionality, which makes the HJI PDEs intractable for problems with large state space.
\rev{There are several papers on computing approximate optimal solutions
for pursuit-evasion games that also bypass the computational intractability of
solving HJI equations explicitly \cite{liu2014evasion,liu2013evasion,takei2014efficient}.}

The explicit policy method is widely used for multi-player PEGs that require scalability in the number of agents.
For scenarios involving multiple pursuers, specific control strategies have been proposed with the analyses on their performance guarantees.
Approaches based on Voronoi tessellation and area minimization can be found in various works \cite{Huang2011a,Pierson2017,Zhou2016}.
A variant of such work proposes a so called relay pursuit to improve the overall efficiency by selecting one pursuer to actively go after the evader \cite{Bakolas2012}, and it has been applied to a more complex scenario \cite{Selvakumar2019}.
A behavior called the cyclic pursuit uses a chain of pursuers to encircle a target \cite{Bopardikar2009,Kim2007a}.
For a non-adversarial scenario where there is no evasive maneuver, the problem is formulated as the vehicle-routing problem~\cite{Agharkar2014}.
Evasive maneuvers have also been consider in the scenario with one pursuer and multiple evaders \cite{Fuchs2010,Scott2018}.

The problem becomes more challenging when there are multiple pursuers and multiple evaders.
The underlying question is ``which pursuer should go after which evader?''
In \cite{Pierson2017}, a Voronoi-tessellation based approach was used to directly obtain the desired direction of motion.
In \cite{Makkapati2019}, a task allocation approach was proposed, where the solution to the multiple pursuers vs.\ one evader problem was used to assign a unique pursuer for each evader so that capture in minimum time is guaranteed.

Specifically for the reach-avoid game played between multiple defenders and multiple attackers, \cite{Chen2014a} approximated the multi-player game as a combination of two-player games.
In contrast to a more conventional PEG that considers the time of capture, we must consider whether the given pursuer can capture the attacker before it reaches the target.
To obtain this feasibility (capturability) information, the solutions to the two player games (strategies and winning regions) were obtained by numerically solving the associated HJI PDE \cite{Chen2014a,chen2017}.
\rev{As an advantage of using a numerical approach, the authors were able to handle complex environments with obstacles.}
These solutions were used to formulate the design of defense policy as an assignment problem.

Following the approach taken in \cite{Chen2014a,chen2017}, this paper starts by identifying the solution to the two-player game: the game played between one defender and one intruder.
Although the two-player game has been solved either numerically \cite{Chen2014a,chen2017}, or under restricted assumptions (circular perimeter or equal speed) \cite{shishika2018cdc}, we analytically solve the problem for arbitrary convex shapes.
This is enabled due to the constraint that the defender moves on the perimeter.

\rev{Our analytical solution has several advantages over the numerical one. First, it lends itself to convenient geometric interpretations such as \emph{approach angle}. In addition, while the numerical approach requires us to compute the solution offline and store the data (i.e., a look up table that gives control inputs from the current positions), analytical approach efficiently computes the control input online, and thus requires much less memory.}


In addition, the derived solution exhibits an interesting contrast to the
solutions based on \emph{dominance region}, which was used in the original work by Isaacs \cite{Isaacs} and also in \cite{Oyler2016}.
The intruder-dominated region contains all the points that the intruder can reach first regardless of the defender's strategy \cite{Oyler2016}.
One can conclude that the intruder can successfully reach the target/perimeter if the intruder-dominated region intersects the target region.
However, our analysis shows that such condition is only sufficient and not necessary in the perimeter defense game.


\if \ARXIVversion1
We also extend the existing assignment method by incorporating a cooperative defense performed by two defenders.
To this end, we analytically solve the game played between two defenders and one intruder.
Then the solution to this two vs.\ one game is incorporated in the extended assignment policy.

The main contributions of the paper are
(i) the solution to the one vs.\ one game;
(ii) the solution to the two vs.\ one game that shows the benefit of defender cooperation; and
(iii) the analysis on the optimality of the derived strategies. 
\rev{These results are essential building blocks to solve the game played between multiple defenders and multiple intruders \cite{shishika2020ral}.
Finally, we also present and discuss three different assignment-based defense policies that utilize the aforementioned results.
}
\fi

\if \ARXIVversion0
We also solve the game played between two defenders and one intruder.
It is shown that the defenders benefit from cooperatively performing a pincer movement, rather than playing one vs.~one games separately.

The main contributions of the paper are
(i) the solution to the one vs.\ one game;
(ii) the solution to the two vs.\ one game; and
(iii) the analysis on the optimality of the derived strategies. 
These results are essential building blocks to solve the game played between multiple defenders and multiple intruders \cite{shishika2020ral}.
\fi

In our previous work \cite{shishika2018cdc} the perimeter-defense game was solved on a circular perimeter with a formulation that is not extensible to general shapes.
This paper uses a formulation that can treat any convex shapes including the circular perimeter.
While the extension to polygonal perimeter was discussed in \cite{shishika2018cdc}, the result was limited to the case where the defender and the intruder have the same speed limits, \rev{which allowed us to simplify the analysis.}
This paper \rev{fills the existing gap by accommodating} a more general case where the defender has any speed that is equal or higher than the intruder.
Finally, the discussion of the payoff functions, for which the derived strategies are optimal, has not been published before.

The paper is organized as follows.
Section~\ref{sec:problem_formulation} formulates the problem.
Section~\ref{sec:twoplayergame} solves the game played by one defender and one intruder.
Section~\ref{sec:threeplayergame} introduces the cooperative aspect by solving the game played by two defender and one intruder.
\if\ARXIVversion1
Section~\ref{sec:team_strategy} proposes the defender team strategy using the results of one vs.\ one and two vs.\ one games.
\fi
Section~\ref{sec:simulation} presents the numerical results.



\section{Problem formulation}\label{sec:problem_formulation}

This section formulates the reach-avoid game for defenders constrained on a perimeter.
The target $\mc{T}\subset \mathbb{R}^2$ is assumed to be a convex region on a plane, and its perimeter is given by an arc-length parameterized curve $\curve: [0,L)\rightarrow \partial \mc{T}$, where $L$ denotes the perimeter length.\footnote{\rev{In case the target is concave, the results of this paper can be applied by taking the convex hull of the original region and by deploying defenders to protect this virtual target region.}}

We use $s\in[0,L)$ to denote the arc-length position on the curve measured in counter-clockwise (ccw) direction.
The tangent vector of the curve at $s$ is denoted by 
\beq
\tangent(s) \triangleq \frac{d \curve(s)}{d s}.
\eeq
For any two points/vectors in $\mathbb{R}^2$ we denote the relative vectors using 
\beq
\relvec{a}{b} \triangleq \mf x_b - \mf x_a,
\eeq
and the unit vectors using $\hat{\mf x} = \frac{\mf x}{\| \mf x \|}$.
The arc-length from point $s_a$ to $s_b$ on the curve in ccw direction is denoted by 
\beq
\arclength{a}{b} \triangleq (s_b - s_a) \text{\;mod\;} L,
\eeq
for example, $\arclength{a}{b}+\arclength{b}{a} = L$.
The segment starting from $s_a$ and ending at $s_b$ in ccw direction is denoted by $\segment{s_a}{s_b} \triangleq \{s_x\,|\,\arclength{a}{x}\leq\arclength{a}{b}\}$.
We use $(s_a,s_b)$ when the endpoints are not included.

%
A set of $N_D$ defenders $\{D_i\}_{i = 1}^{N_D}$ are constrained to move on the perimeter.\footnote{\rev{This assumption is motivated by various scenarios, for example, in which ground defenders are deployed to protect a building that they cannot move through.}}
The position of the $i$th defender is described by $s_{D_i}$ or $\mf x_{D_i} = \curve(s_{D_i})$.
The defender's control input is the signed speed: $\dot{s}_{D_i} = \omega_{D_i}$ or $\dot{\mf x}_{D_i} = \controld_{D_i}\tangent(s_{D_i})$ with the constraint $|\controld_{D_i}|\leq 1$.

A set of $N_A$ intruders $\{A_j\}_{j = 1}^{N_A}$ have first-order integrator dynamics in $\mathbb{R}^2$.
The control inputs are the velocities; $\dot{\mf{x}}_{A_i} = \mf u_{A_i}$ with the constraint $\| \mf u_{A_i} \| \leq \nu$.
It is assumed that the defender is \rev{at least as fast as the intruder}:\footnote{\rev{The case with faster intruders (i.e., $\nu>1$) requires a separate treatment and is a subject of ongoing work. It is easy to see that if $\nu>1$, and also if the capture is defined by zero distance, then the intruder can always win.}}
\bql
\nu  \in(0,1].
\eql
\rev{This is a generalization of the case with $\nu=1$ studied in \cite{shishika2018cdc} and removes some simplifications (see Sec.~\ref{sec:special_case}).}

We assume that each player has access to the current state and the speed ratio $\nu$.
\rev{However, the players do not know the instantaneous control action of the opponent.}

In a microscopic view, an intruder $A_i$ scores if it reaches the target ($\mf x_{A_i}\in \partial \target$) without being captured by the defenders.
We use zero distance to define capture: i.e., $\|\mf x_A - \mf x_D\|=0$,
\rev{however, the extension to the case with non-zero capture radius is also straightforward \cite{shishika2019team}.}
The defender moves on the perimeter to either intercept the intruder or prevent it from scoring indefinitely.

\if\ARXIVversion0
As the building blocks to analyze the multi-player game, we solve the game played by one defender and one intruder, and also by two defenders and one intruder.
\paragraph*{Problem:}
\fi

\if\ARXIVversion1
As the building blocks to analyze the multi-player game, we solve the game played by one defender and one intruder, and also by two defenders and one intruder.
\paragraph*{Problem 1:}
\fi
Find the \rev{\emph{barrier} surface \cite{Isaacs}} that divides the state space into intruder-winning  and defender-winning configurations.
In each region, what are the strategies to be used by the players?

\if \ARXIVversion0
In a macroscopic view, let $Q\in\mathbb{N}$ denote the number of intruders that reach the perimeter.
The intruder team maximizes $Q$ while the defender team minimizes it.
The team strategies for this macroscopic problem are omitted due to page constraints, and is available in the unabridged version \cite{shape_long}.
\fi

\if \ARXIVversion1
In a macroscopic view, let $Q\in\mathbb{N}$ denote the number of intruders that reach the perimeter.
The intruder team maximizes $Q$ while the defender team minimizes it.
\paragraph*{Problem 2:}
Given an initial configuration of the game and the speed ratio $\nu$, \rev{what are the upper and lower bounds on the score $Q$, and what are the associated team strategies to ensure that bound?}

We address these problems in the following sections.
\fi

\if \ARXIVversion0

\fi



\section{One vs.\ One Game} \label{sec:twoplayergame}
This section solves the game played between one defender and one intruder.
The states of the system are $[s_D,\xa]$ and the dynamics are $[\dot{s}_D,\dot{\mf x}_A]= [\omega_D,\mf u_A]$.
The terminal surface corresponding to intruder's win is $\{[s_D,\xa]\,|\,\xa \in \target$ and $\|\xa - \curve{s_D} \|>0\}$.
The terminal condition for defender's win is discussed later in Sec.~\ref{sec:winning_regions}.

We first introduce some relevant geometries, and then solve the \emph{game of kind} to find the \emph{barrier} surface \cite{Isaacs} that divides the game space into the intruder-winning and the defender-winning regions.
We also discuss the objective functions for which the derived strategies are also optimal in the \emph{game of degree}.

\subsection{Geometries}
\begin{figure}
\begin{center}
\includegraphics[width=.49\textwidth]
{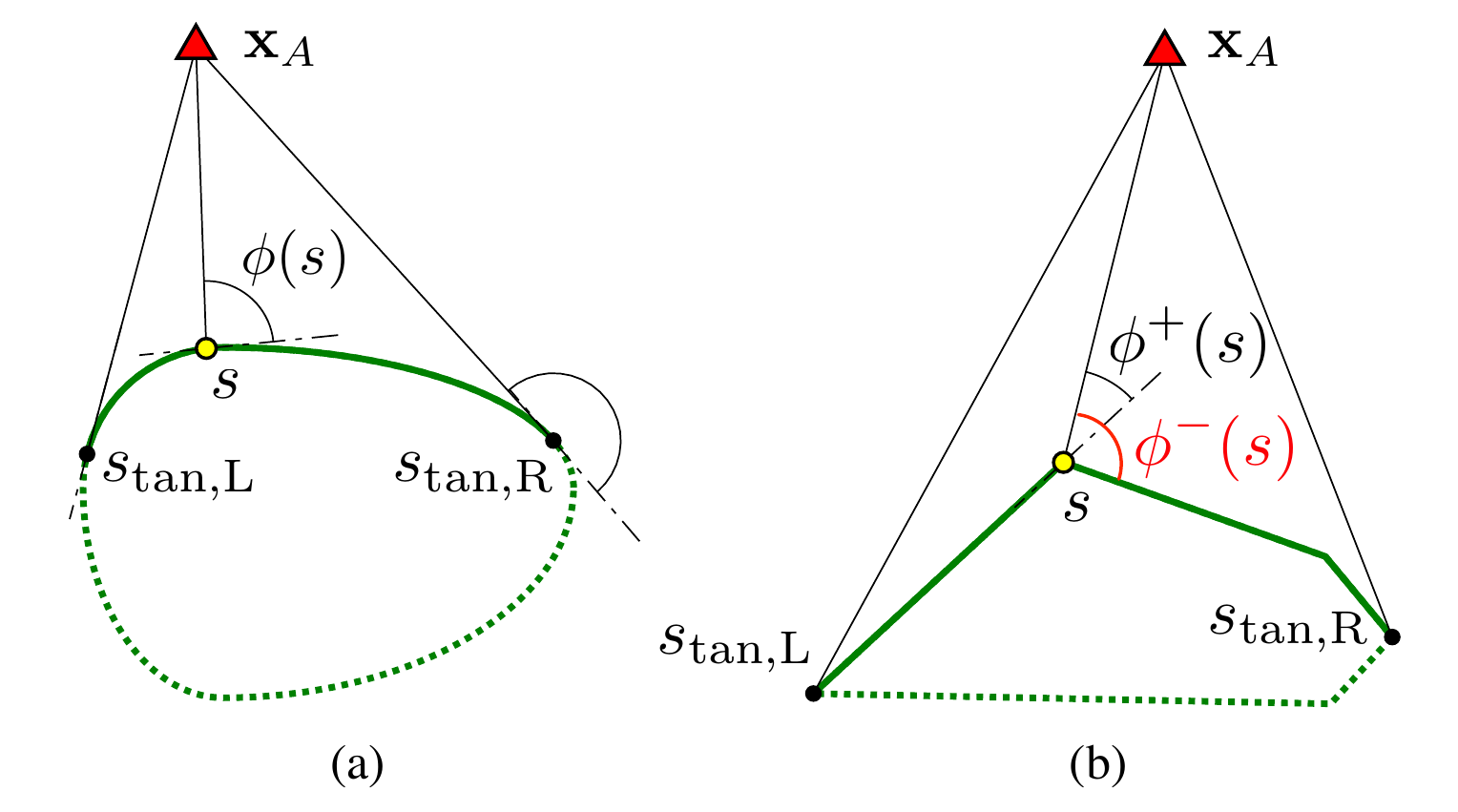}    
\caption{Illustration of the tangent points and the approach angle.
The segment $\sdirect$ is indicated with the solid line.
(a) A continuously differentiable perimeter.
(b) A polygonal perimeter.}  
\label{fig:approach_angle}                                 
\end{center}                                 
\end{figure}

Let $\stanr$ and $\stanl$ denote the points where the tangent lines from $\mf x_A$ touch the perimeter (see Fig.~\ref{fig:approach_angle}a).
Considering the directions from the perspective of a defender facing outward from the perimeter, the subscript $_\text{R}$ corresponds to the ``right'' or clockwise (cw) direction of motion, and $_\text{L}$ corresponds to the ``left'' or counter-clockwise (ccw).
We use 
\beq
\sdirect(\xa) \triangleq \segment{\stanr}{\stanl}
\eeq
to denote all the points on the perimeter that the intruder can reach by a straight-line path.
Note that these geometries are independent of the defender position.

For a given point $s_B\in \sdirect$ and \rev{$s_D\notin\sdirect$} consider the following quantity\footnote{\rev{The restriction $s_D\notin\sdirect$ will be removed after Remark~1.}}:
\bnl
J_L(s_B;s_D,\xa) \triangleq \arclength{D}{B} - \frac{\| \gamma(s_B) - \mf x_A \|}{\nu}.
\label{eq:JL}
\enl
The first term is the ccw distance from the defender to $s_B$, and $\|\gamma(s_B)-\xa\|$ is the distance from the intruder to $s_B$.
Hence, recalling that the defender and the intruder has the speed 1 and $\nu$ respectively, $J_L$ describes how much longer it takes for the defender to reach $s_B$ than it takes for the intruder, when the defender moves ccw and the intruder moves on a straight line path towards $s_B$.
The subscript $_L$ is used to highlight that we assume the engagement in the ``left'' or ccw direction.

Suppose the game starts at $t=0$ and the intruder reaches $s_B$ at time $t_F$ before the defender does.
Then $\arclength{D}{B}(t_F)=\arclength{D}{B}(0)-t_F\omega_D\geq \arclength{D}{B}(0)-t_F = \arclength{D}{B}(0) - \frac{\| \gamma(s_B) - \mf x_A(0) \|}{\nu}$.
Therefore, a positive $J_L(s_B)$ can also be interpreted as the expected arc-length distance between the intruder and the defender when the intruder reaches $s_B$.
\rev{To focus on the geometry, we defer the question of defender's optimal direction of motion, and whether the intruder should employ a straight line path or not, to the later sections (e.g., Remark~\ref{rem:straight_line_path}).}



Restricting ourselves to straight line paths for now, the intruder maximizes $J_L$ by finding the optimal breaching point $s_B$.
The derivative is given by
\bqn
\frac{d J_L}{d s_B} &=& \frac{d}{d s_B} (s_B - s_D) - \frac{1}{\nu}\frac{d}{d \curve}\|\curve-\mf x_A\|  \cdot \frac{d \curve(s_B)}{d s_B}\\
&=& 1 - \frac{1}{\nu}\frac{\curve(s_B)-\mf x_A}{\| \curve(s_B)-\mf x_A \|}\cdot \tangent(s_B),\\
\eqn
where the dot product in the second term is related to the \emph{approach angle} defined in the following:
\begin{definition}
Suppose the intruder position $\mf x_A$ is given.
Then for $s \in\sdirect$, we define the {\bf approach angle} to be
\bql
\phi(s) \triangleq \cos^{-1} \left( \frac{\bs \gamma(s) - \mf x_A}{\| \bs \gamma(s) - \mf x_A\|} \cdot \bs T(s)\right)\in [0,\pi].
\eql
For a perimeter with discontinuous tangent vector (e.g., polygonal perimeter), we use $\phi^-(s)$ and $\phi^+(s)$ to denote the approach angles before and after the discontinuity (in ccw direction).
\end{definition}
Note that $\phi$ is non-increasing in ccw direction due to the convexity of $\target$,\footnote{\rev{One can easily verify this by observing that $\tangent(s)$ and $\curve(s)-\xa$ rotate in ccw and cw direction respectively for increasing $s\in\sdirect$.}} and for a \rev{continuously differentiable} perimeter, we always have $\phi(\stanr)=\pi$ and $\phi(\stanl)=0$ (see Fig.~\ref{fig:approach_angle}b).

Using the approach angle, the derivative is described as:
\beq
\frac{d J_L}{d s_B} = 1 - \frac{\cos \phi(s_B)}{\nu},
\eeq
which gives the following result:
\bnl
\frac{d J_L}{d s_B} = \left\{ \begin{array}{l l}
positive & \text{if\;} \phi(s_B) > \phi_L^*\\
0           & \text{if\;} \phi(s_B) = \phi_L^* \\
negative & \text{otherwise},
\end{array}\right.
\label{eq:partial_JL}
\enl
where 
\bql
\phi_L^*=\acos(\nu).
\eql
This result provides the critical breaching point that maximizes $J_L$ as follows:
\begin{definition}
We define {\bf left breaching point} $ s_L(\xa) \in \sdirect$ to be the point that maximizes $J_L$.
For a \rev{continuously} differentiable $\curve(s)$, it is the unique solution of $\phi(s) = \phi_L^*$, i.e., 
\emph{
\bnl
s_L(\xa) = \phi^{-1}\left( \acos\nu \right).
\enl
}
For a perimeter \rev{with discontinuous tangent vector (e.g., polygonal perimeter)}, $s_L(\xa)$ is a unique point that satisfies either of the following conditions:
\bnl
\left\{ \begin{array}{l l}
\phi(s)= \phi_L^* & \text{($s_L$ is on a continuous part),}\\
\phi(s)^+  < \phi_L^*< \phi^-(s)  & \text{($s_L$ is on a vertex)}.
\end{array}
\right.
\label{eq:sL_poly}
\enl
\end{definition}
Due to the monotonicity of $\phi(s)$ on a convex perimeter, $s_L$ is always unique, and it can be found by a simple search on a one-dimensional space.
Note also that $s_L$ is obtained analytically for some special cases discussed in Sec.~\ref{sec:special_case}.

\begin{remark}[Limiting cases]
\label{rem:extreme_cases}
If $\nu=1$, then we always have $s_L=\stanl$, because $\phi_L^*=0$ and $\phi(\stanl)=0$.
When $\nu\rightarrow0$ the optimal approach angle becomes $\phi_L^*\rightarrow \frac{\pi}{2}$, in which case $s_L$ is equivalent to the closest point on the perimeter from $\xa$.
\end{remark}

\rev{Now we consider all defender locations by removing the restriction $s_D\in\sdirect$. With this extension, the left breaching point $s_L$ does not maximize $J_L$ if $s_D \in [\stanr,s_L]$, however, we will show in Sec.~\ref{sec:winning_regions} that $s_L$ and its counterpart $s_R$ are the only points necessary in defining the optimal strategies.}

\begin{figure}[t]
\begin{center}
\includegraphics[width=.49\textwidth]
{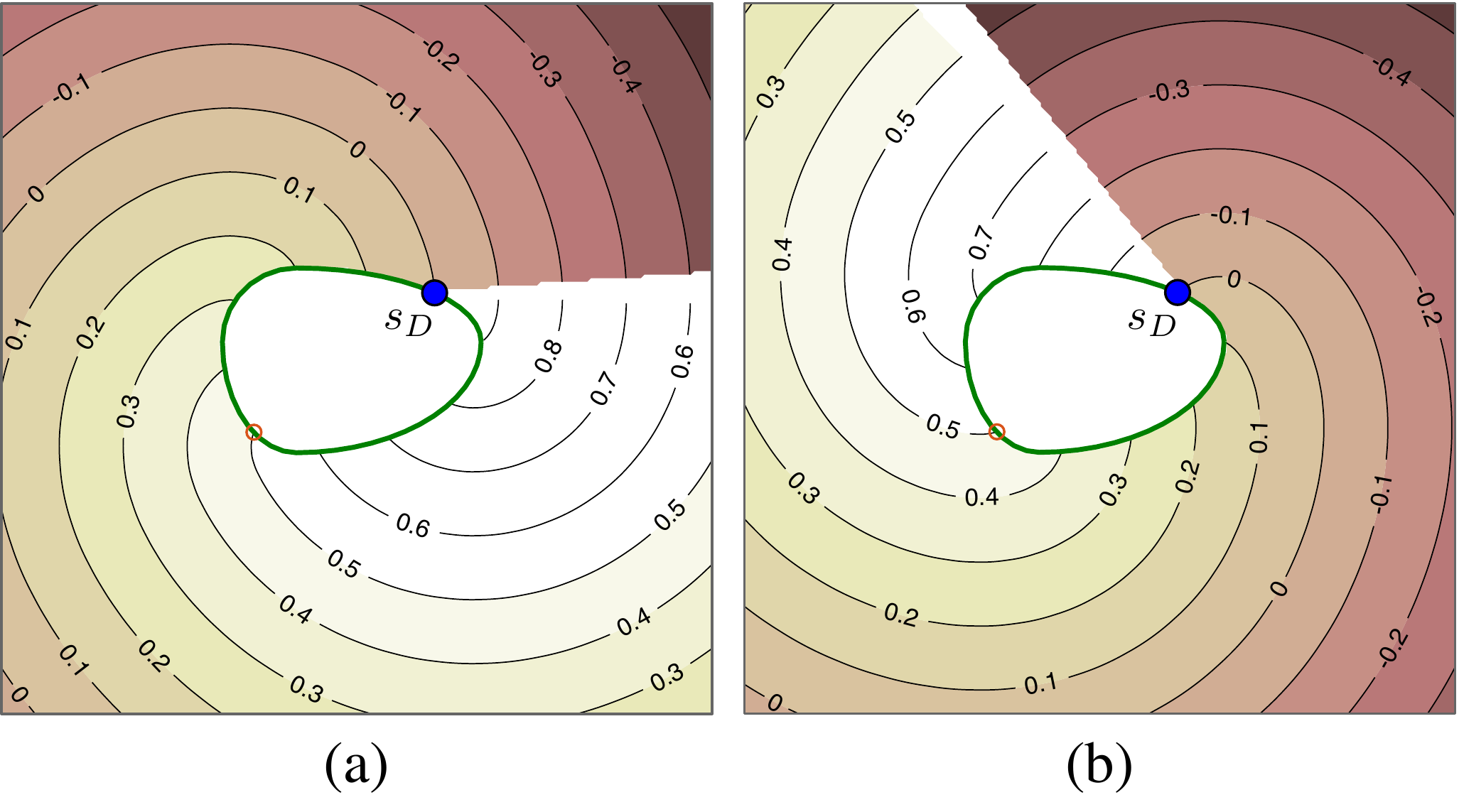}    
\caption{Level sets of $J_L^*(s_D,\xa)$ (left) and $J_R^*(s_D,\xa)$ (right) for a specific value of $s_D$ ($\nu=0.9$).
}  
\label{fig:levelcurve_LR}                                 
\end{center}                                 
\end{figure}

For given positions $s_D$ and $\xa$, we define the following function that gives the critical value of $J_L$:
\bnl
J_L^*(s_D,\xa)  \triangleq J_L(s_L) = \arclength{D}{L} - \frac{\| \gamma(s_L) - \mf x_A \|}{\nu}.
\label{eq:left_objective}
\enl
\rev{Figure~\ref{fig:levelcurve_LR}a shows the level sets of $J_L^*$ for a specific value of $s_D$. The discontinuity corresponds to the manifold where $s_L(\xa)=s_D$.}

For a similar analysis on the cw motion of the defender, consider the following function:
\bnl
J_R(s_B;s_D,\xa) = \arclength{B}{D} - \frac{\| \mf x_A - \curve(s_B)  \|}{\nu},
\label{eq:JR}
\enl
where the arc-length computation is now $\arclength{B}{D}$.
With the same process, 
\rev{we define the \emph{right breaching point}, $s_R$, to be the solution to}
\bnl
\phi(s_R) = \phi^*_R = \pi-\acos(\nu).
\enl
We define a function for the critical value as
\bnl
J_R^*(s_D,\xa)  \triangleq  J_R(s_R) = \arclength{R}{D} - \frac{\| \mf x_A - \curve(s_R)  \|}{\nu}.
\enl

\begin{figure}[t]
\begin{center}
\includegraphics[width=.48\textwidth]
{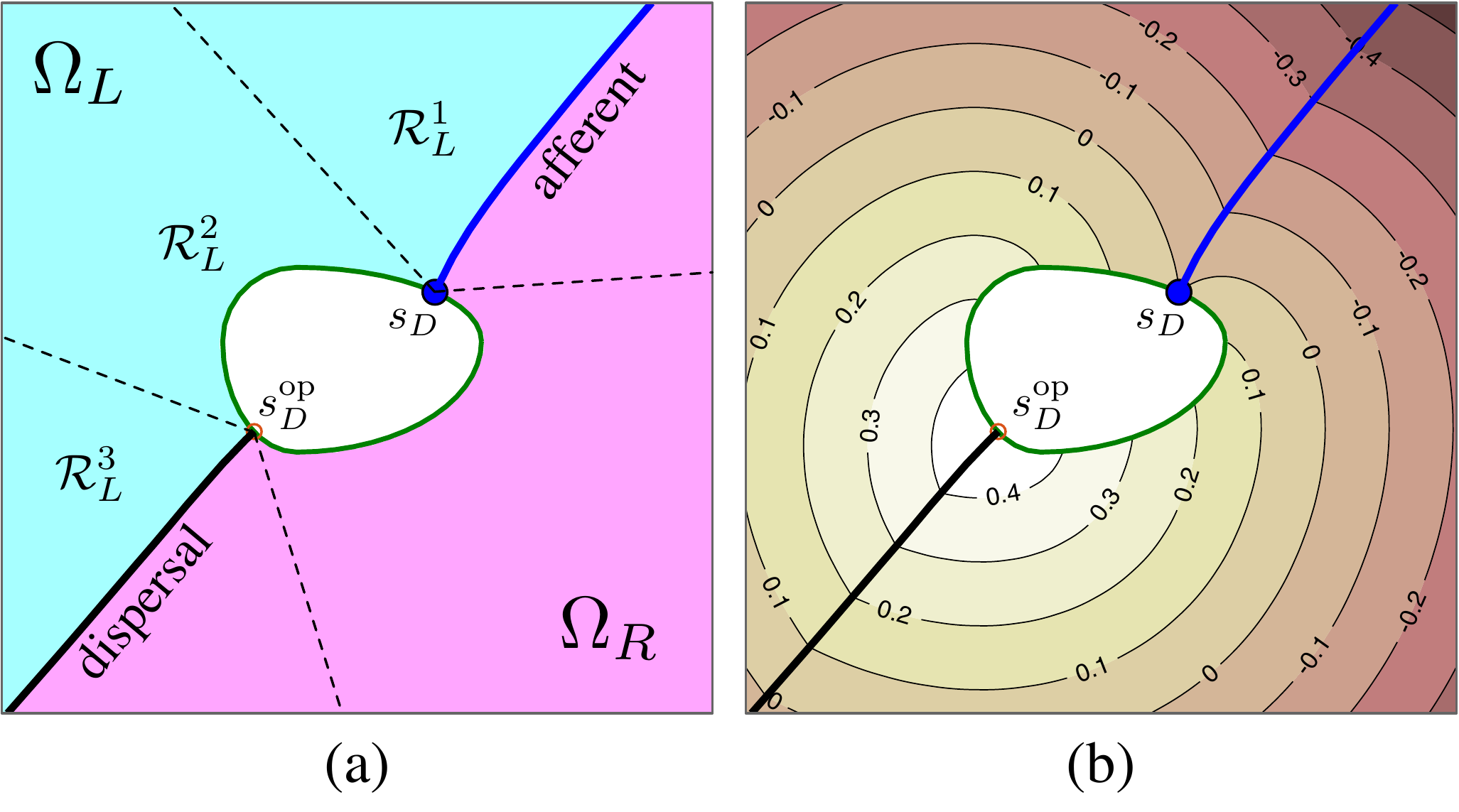}    
\caption{
Singular surfaces for $\nu=0.9$.
(a) Left region (cyan) and right region (magenta).
The left region is further partitioned into three regions.
(b) Level sets of $V$.
}  
\label{fig:dispersal_surface}
\end{center}
\end{figure}

Next we use the two functions $J_L^*$ and $J_R^*$ to divide the game space into ``right side'' and ``left side'' with respect to the position of the defender.
Let $\sDoppo$ be the farthest (opposite) point from the defender on the perimeter.
The partitioning will be given by the singular surface defined in the following:
\begin{definition}
Consider the surfaces defined by
\bnl
\singsurface(s_D) = \{\mf x_A \;|\; J_L^*(\mf x_A, s_D)=J_R^* (\mf x_A, s_D)\}.
\enl
The one extending from $s_D$ is called the \textbf{afferent surface}, $\affsurface$, and the other extending from $\sDoppo$ is called the \textbf{dispersal surface}, $\dispsurface$ (see Fig.~\ref{fig:dispersal_surface}a) \cite{Isaacs}.
\end{definition}
The singular surfaces are defined in the three-dimensional state space, but for convenience, we  look at the ``two-dimensional slice'' by considering a specific value of $s_D$.
The singular surfaces divide the entire game space into two regions.
We define them as the \textbf{left region}, $\Omega_L(s_D)$, and the \textbf{right region}, $\Omega_R(s_D)$ (see Fig.~\ref{fig:dispersal_surface}a).
\revrev{As one can see from the definition, there will be two equally good strategies when the states are on the singular surface. However, we will later show that such non-uniqueness does not prevent us from identifying the barrier.}

Let $\sleft=\segment{s_D}{\sDoppo}$ and $\sright=\segment{\sDoppo}{s_D}$ denote the segments of the perimeter to the left and right of the defender.
Whether the intruder is in the left region or not can be tested using the location of the breaching points ($s_L$ and $s_R$), and the relation between the values $J_L^*$ and $J_R^*$.
If $\xa \in \Omega_L(s_D)$, then $\xa$ is in one of the following three regions (see Fig.~\ref{fig:dispersal_surface}a):
\bnl
\Rleft^1 &=& \{\xa \,|\, s_L \in \sleft, s_R \in \sright, J_L^* > J_R^*\} \notag\\
\Rleft^2 &=& \{\xa \,|\, s_L \in \sleft, s_R \notin \sright \} \\
\Rleft^3 &=& \{\xa \,|\, s_L \notin \sleft, s_R \notin \sright, J_L^* < J_R^*\}.\notag
\label{eq:regions1v1}
\enl
If the states [$s_D,\xa$] satisfy none of the above three conditions, and if $J_L^*\neq J_R^*$, then we have $\xa \in \Omega_R(s_D)$.

Finally, we merge the two objective functions as follows:
\bnl
V(\mf x_A, s_D)= \left\{ \begin{array}{l l}
J_L^*(\mf x_A, s_D) & \text{\;\;if\;\;} \mf x_A \in \Omega_L(s_D)\\
J_R^*(\mf x_A, s_D) & \text{\;\;otherwise}.
\end{array}\right.
\label{eq:value}
\enl
Fig.~\ref{fig:dispersal_surface}b shows the level sets of $V(s_D,\xa)$.\footnote{The evolution of these level sets with the defender position is illustrated in:  \texttt{https://youtu.be/h0{\char`_}VqJbNsQc}}
We later show in Sec.~\ref{sec:optimality1v1} that this is the \emph{value of the game} for some payoff functions.
\if \ARXIVversion0
{\color{\maincolor}
Also, an algorithmic description for the computation of $s_L$ is provided in the unabridged version \cite{shape_long}.
}
\fi

\if\ARXIVversion1
We close this section by providing an algorithm to compute $s_L$ for general $\nu\in(0,1)$. Note that in the special case where $\nu=1$ or $\nu\rightarrow 0$, $s_L$ is immediately obtained as discussed in Remark~\ref{rem:extreme_cases}.
 \begin{algorithm}[h]
 \caption{Finding left breaching point $s_L$ \label{alg:sL}}
 \begin{small}
 \begin{algorithmic}[1]
\State {\bf Input}: $\xa$, $\curve$, and $\nu$
\State Compute tangent points $\stanl$ and $\stanr$
\State $\sdirect\gets\segment{\stanr}{\stanl}$
\If{$\exists$ a vertex $s\in\sdirect$ s.t. \eqref{eq:sL_poly} is true}
    \State{$s_L \gets s$}
\Else
    \State{$s_L \gets \argmin_{s\in\sdirect} |\phi(s)-\cos^{-1}(\nu)|$}
\EndIf
\State {\bf Return}: $s_L$ 
\end{algorithmic}
\end{small}
\end{algorithm}

The condition in line~4 is only necessary for perimeters that are non-differentiable. 
It is sufficient to visit non-differentiable vertices in the interval $\sdirect$, and test the condition \eqref{eq:sL_poly}.
When there is no such critical point, then the optimization in line~7 is performed.
The simplest way to perform this optimization is to discretize the interval $\sdirect$ into a finite set of points and evaluate the right-hand side, which is practically fine since the complexity of the search grows only linearly with the resolution.
To improve the efficiency, one can also use, for example, the bisection method \cite{corliss1977root}.

The right breaching point $s_R$ can be computed in a similar way.
Once these breaching points are found, $J_L^*$ and $J_R^*$ are immediately obtained using \eqref{eq:JL} and \eqref{eq:JR}.
\fi

\subsection{Winning Regions} \label{sec:winning_regions}
\begin{figure}[t]
\begin{center}
\includegraphics[width=.4\textwidth]
{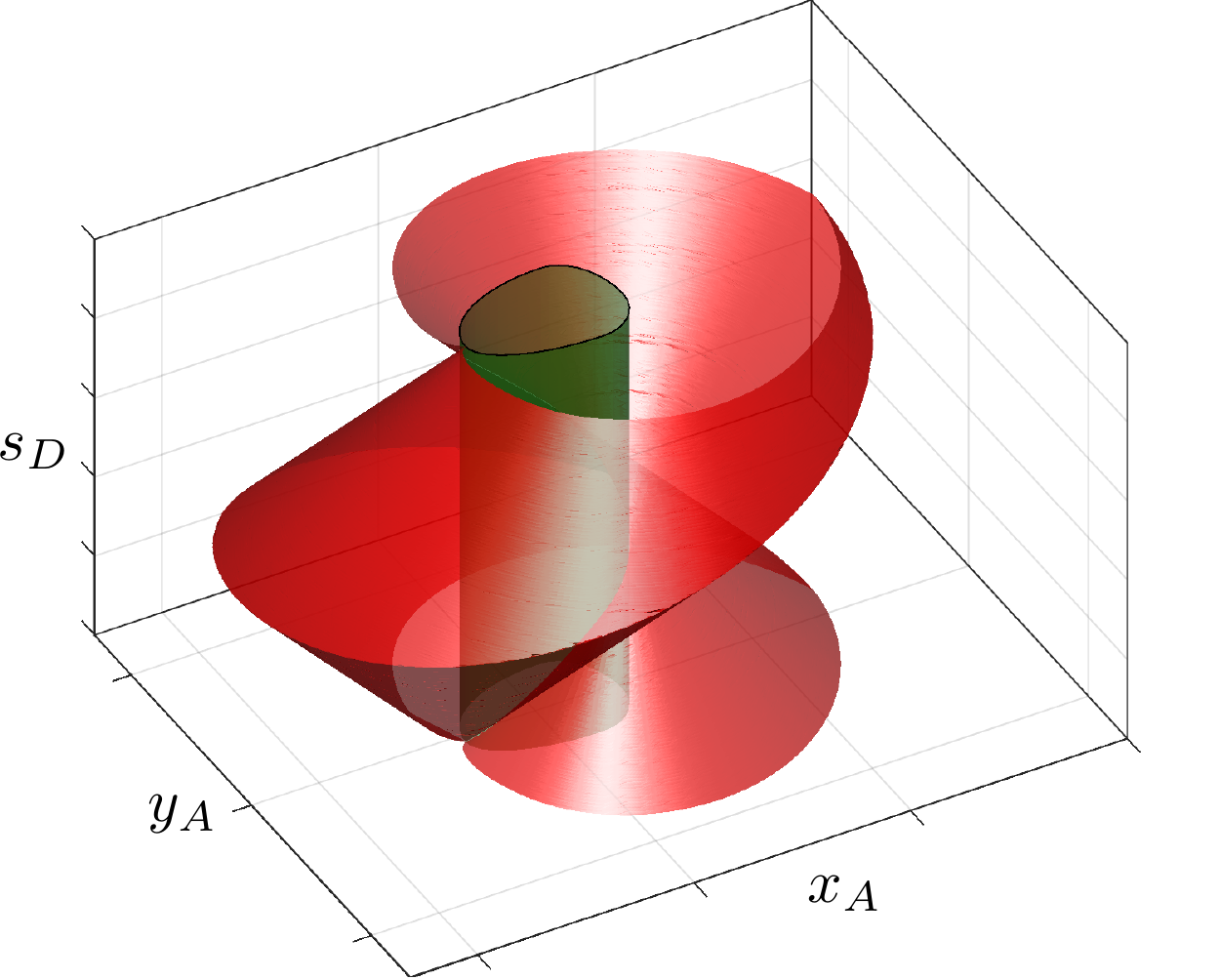}    
\caption{
The barrier surface (depicted in red).
The green cylinder depicts the perimeter shape extruded vertically.
Intruder winning region is the interior of the barrier surface.
}  
\label{fig:winning_region_3d}                                 
\end{center}                                 
\end{figure}

This section proves that the barrier for the game of kind is given by the zero level set of $V$ defined in \eqref{eq:value}.
Fig.~\ref{fig:winning_region_3d} depicts the surface $V(s_D,\xa)=0$ in the three dimensional state space.
For convenience, we perform our analysis using the two-dimensional slice at $s_D$ corresponding to the location of the defender.
We define the intruder winning region as
\bnl
\awinr(s_D) = \{\mf x_A\,|\,V(s_D,\xa)>0\}.
\enl
We first show that the intruder can guarantee its victory if it starts inside $\awinr$.

\begin{lemma}\label{lem:intruder_winning_region}
If the initial configuration is such that $\xa\in\awinr(s_D)$ (i.e., $V>0$), then regardless of the defender strategy, the intruder guarantees its win using the following feedback strategy:
\bnl
\mf u_A^*= \left\{ \begin{array}{l l}
\nu \unitvec{A}{L} & \text{\;\;if\;\;} \mf x_A \in \Omega_L(s_D)\\
\nu \unitvec{A}{R}& \text{\;\;otherwise},
\end{array}\right.
\label{eq:ua_star}
\enl
\rev{where $\unitvec{A}{L} = \frac{\curve(s_L)-\xa}{\|\curve(s_L)-\xa\|}$, and $\unitvec{A}{R} = \frac{\curve(s_R)-\xa}{\|\curve(s_R)-\xa\|}$.}
\footnote{\revrev{For conciseness, we take the convention that the intruder treats the singular surface as part of $\Omega_R$. On the singular surface, the two actions in \eqref{eq:ua_star} are equally good, and this choice is inconsequential towards the outcome of the game.}}
\end{lemma}
\begin{pf}
Suppose $\mf x_A\in\Omega_L(s_D)$ without the loss of generality.
We consider two cases: (i) $s_L\in \segment{s_D}{\sDoppo}$, and (ii) $s_L\in \segment{\sDoppo}{s_D}$.
In either case, we know that the intruder reaches $s_L$ first if the defender moves ccw, because $J_L^* = V > 0$.

In the first case when $s_L\in \segment{s_D}{\sDoppo}$, it is clear that the cw motion by the defender takes longer time to reach $s_L$ than the ccw motion since $\arclength{L}{D}>\arclength{D}{L}$.
Therefore, the intruder can reach $s_L$ first regardless of the defender strategy.
The set of all intruder positions corresponding to the first case is shown as the shaded (cyan) region in Fig.~\ref{fig:winning_region_partial}.
\begin{figure}[t]
\begin{center}
\includegraphics[width=.44\textwidth]
{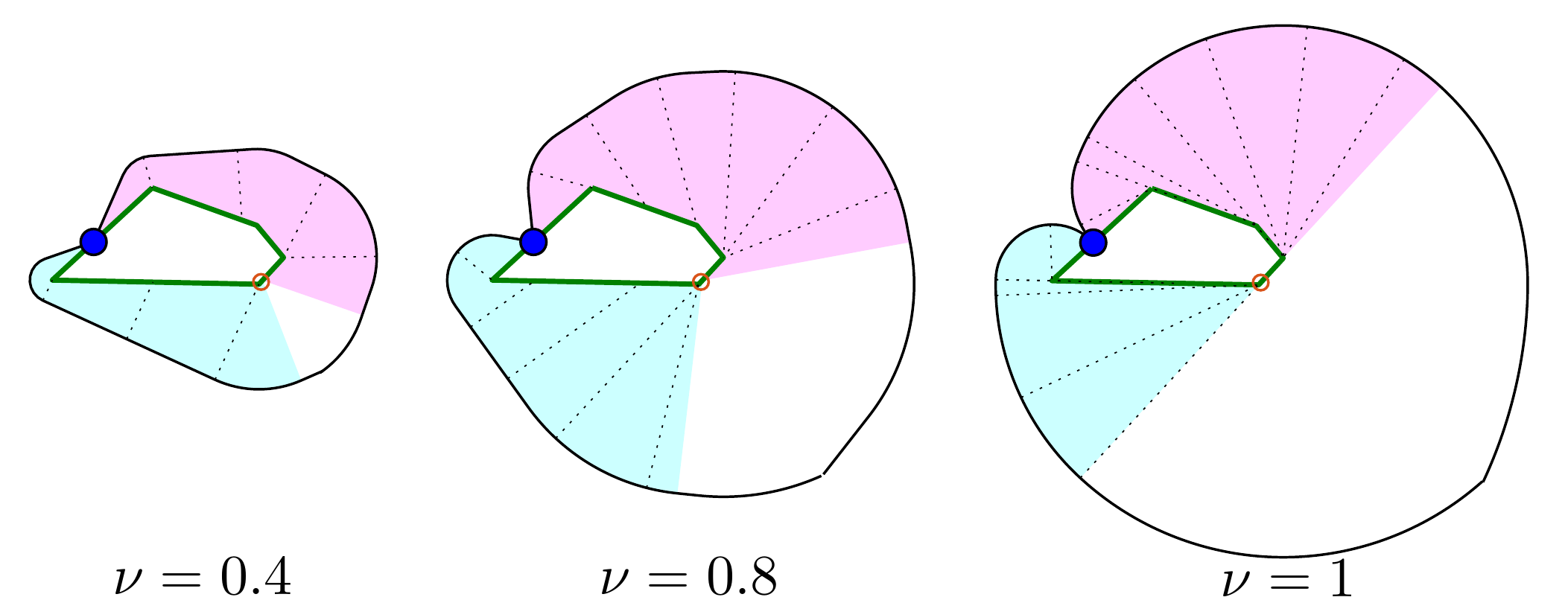}    
\caption{Intruder-winning region under the constraint $s_L \in \mc S_L$ (cyan) and $s_R \in \mc S_R$ (magenta), for varied intruder speed $\nu$.
The dotted lines illustrate the corresponding intruder paths.
}  
\label{fig:winning_region_partial}                                 
\end{center}                                 
\end{figure}
%

\begin{figure}[t]
\begin{center}
\includegraphics[width=.49\textwidth]
{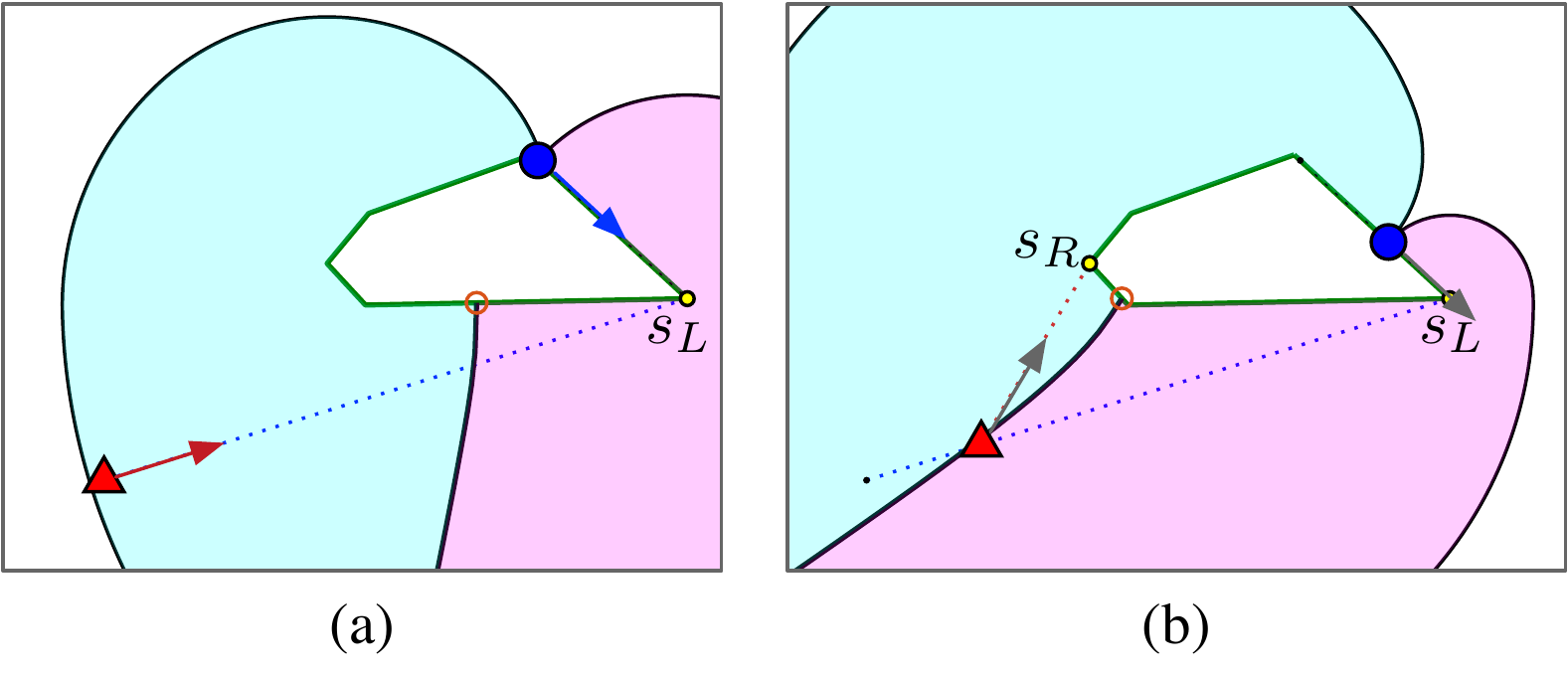}    
\caption{
Engagement when the game starts in a configuration with $s_L\notin\sleft$ and $s_R\notin\sright$.
(a) Defender takes a suboptimal strategy aiming at $s_L$.
(b) Intruder enters $\Omega_R$ and switches its heading to $s_R$.
}  
\label{fig:intruder_winning}                                 
\end{center}                                 
\end{figure}
The second case where $s_L\in\segment{\sDoppo}{s_D}$ (corresponding to the white region in 
Fig.~\ref{fig:winning_region_partial}) is more subtle since \revrev{the defender may be tempted to move cw} to reach $s_L$ before the intruder does (e.g., see Fig.~\ref{fig:intruder_winning}a).
Suppose the defender takes this strategy: $\omega_D=-1$ (cw motion).
Then $J_L^*$ increases because $\arclength{D}{L}$ in \eqref{eq:left_objective} increases.
Now, there exists a time $t_1$ when $\xa(t_1)\in\dispsurface(s_D(t_1))$, at which point we have
\bql
V(t_1)=J_L^*(t_1)=J_R^*(t_1)>J_L^*(t_0)=V(t_0)>0.
\eql
If the defender continues in cw direction, the intruder enters $\Omega_R$, and the strategy \eqref{eq:ua_star} switches the breaching point to $s_R$.\footnote{The simulation video at {\texttt{https://youtu.be/h0{\char`_}VqJbNsQc}} illustrates the engagement.}
The intruder will reach $s_R$ first because $J_R^*(t_1)>0$ (Fig.~\ref{fig:intruder_winning}b).
If the defender goes back to ccw motion, the intruder stays in $\Omega_L$ and continues towards $s_L$.
The intruder will reach $s_L$ first because $J_L^*(t_1)>0$.

Therefore, no matter what decision the defender makes at this point in time, $V(s_D,\xa)$ stays positive throughout the rest of the game, and the intruder never leaves $\awinr(s_D)$ until it reaches the perimeter. 
Note that, for the defender, this configuration at $t_1$ is strictly ``worse'' than the initial one in the sense that $V$ is now strictly larger than what it was at $t_0$.

\revrev{Also note if $\xa\in\dispsurface$, and if the defender continues to switch its heading 
(according to $\omega_D=-1$ if $\xa\in\Omega_L$, and $\omega_D=1$ otherwise, 
which is the opposite of \eqref{eq:omega_star}),
there will be a ``chatter'' due to infinitely frequent switching in the heading.\footnote{
\revrev{Such phenomena often arise in differential games \cite{von2020robust,pachter2019singular}.
The solution to a differential equation with discontinuous right-hand side (due to chattering) can be provided in the sense of Filippov \cite{filippov2013}.}
}
The intruder will oscillate about $\dispsurface$, and since it always has a velocity component towards the perimeter (due to its convexity), the intruder slide along $\dispsurface$ to approach the perimeter and eventually reach $\sDoppo$. Notice that the defender gains no advantage in the azimuthal proximity to the intruder, and thus a rational defender will in fact never use such a strategy.}
$\blacksquare$
\end{pf}

\begin{remark}[Dominance region]\label{rem:dominance}
For the configuration in Fig.~\ref{fig:intruder_winning}a, 
the analysis based on the dominance region \cite{Oyler2016} will not conclude that the intruder can win the game, because $s_L$ is not in the intruder-dominated region; i.e., the defender has a way to reach $s_L$ before the intruder.
Nevertheless, we have shown that the intruder can win the game by employing a feedback strategy \eqref{eq:ua_star}.
\end{remark}

The result mentioned in Remark~\ref{rem:dominance} is a consequence of the following points: (i) the perimeter acts as an obstacle, and (ii) the defender is protecting a region (and not a single point).
\rev{Rather than moving towards the optimal breaching point in the shortest path, the defender must maneuver so that it does not generate a breaching point that is ``worse'' (corresponding to a higher $V$), as was illustrated in Fig.~\ref{fig:intruder_winning}.
This is why the defender must travel a distance longer than $L/2$ (half of the perimeter length) when the intruder is in the unshaded region in Fig.~\ref{fig:winning_region_partial}}.

\begin{remark}[Straight line path]\label{rem:straight_line_path}
\rev{Consider the case $\xa\in\Omega_L$. Noting that $s_L$ remains constant if $\acontrol=\nu \unitvec{A}{L}$, and that $s_L$ is independent of $s_D$, the strategy in \eqref{eq:ua_star} clearly results in a straight line path towards $s_L$.
Even if the defender behaves suboptimally, as exemplified in Fig.~\ref{fig:intruder_winning}, the intruder's path will still remain piece-wise linear.
This observation combined with the results of Sec.~\ref{sec:optimality1v1} justifies the restriction of the intruder strategy to a set of straight-line paths.
}
\end{remark}

Lemma~\ref{lem:intruder_winning_region} only gives a sufficient condition for the intruder to win.
To prove that it is also a necessary condition, we show that the defender wins if the game starts in a configuration $\xa\notin\awinr(s_D)$.


Recall that the defender wins the game by either intercepting the intruder or preventing it from reaching the perimeter indefinitely.
Related to the latter scenario, we show that the defender is able to stabilize the system around the configuration $\xa\in\affsurface (s_D)$.\footnote{This stabilization is also demonstrated in the simulation video available at { \texttt{https://youtu.be/h0{\char`_}VqJbNsQc}}}
\begin{lemma}\label{lem:stability_around_affsurface}
When $\xa(t_0) \in \affsurface(s_D(t_0))$, then for any intruder control strategy, the defender can maintain the condition $\xa(t)\in\affsurface(s_D(t))$ for all $t>t_0$ using the following control:
\bnl
\omega_D^*(s_D,\mf x_A)= \left\{ \begin{array}{r l}
1 & \text{\;\;if\;\;} \mf x_A \in \Omega_L(s_D)\\
-1& \text{\;\;otherwise}.
\end{array}\right.
\label{eq:omega_star}
\enl
\end{lemma}
\begin{pf}
In the neighborhood of the surface $\affsurface(s_D)$,
consider the error function
$e = J_L^* -  J_R^*$.
Noting that $e>0$ if $\xa\in\Omega_L(s_D)$, and $e<0$ otherwise,
we can rewrite the control as 
$\omega_D^*=\text{sgn}(e)$.
(Note, this expression of control is only valid in the neighborhood of $\xa\in\affsurface(s_D)$.)
The time derivative of the squared error is given by $\frac{d}{dt}e^2=2e(\dot{J}_L^*-\dot{J}_R^*)$, where $\dot{J}_L^*$ is
\bnl
\frac{dJ_L^*}{dt} &=& \dot{s}_L - \dot{s}_D - \frac{\unitvec{A}{L}}{\nu}\cdot \left(\dot{s}_L\tangent(s_L)-\acontrol \right) \notag\\
&=& \dot{s}_L\left( 1-\frac{\cos\phi(s_L)}{\nu} \right) + \frac{\unitvec{A}{L}}{\nu}\cdot \acontrol - \omega_D\notag\\
&=&\frac{\unitvec{A}{L}}{\nu}\cdot \acontrol - \omega_D.
\label{eq:JLdot}
\enl
From the second to the third line, we used the fact that $\dot{s}_L\left( 1-\frac{\cos\phi(s_L)}{\nu} \right)=0$, which we prove in the following.
Observe that a small displacement in $\xa$ moves $s_L$ if it is on a \rev{continuously differentiable} part of the perimeter, but $s_L$ will remain stationary if it is on a vertex (see \eqref{eq:sL_poly}).
When $s_L$ is on a \rev{continuously differentiable} part, we have $\phi(s_L)=\phi_L^*=\acos\nu$, which gives $1-\frac{\cos\phi(s_L)}{\nu}=0$.
When $s_L$ is on a vertex and not moving, we have $\dot{s}_L=0$.

With a similar computation on $\dot{J}_R^*$, the time derivative of the squared error is
\bqn
\frac{\nu}{2}\frac{d}{dt}e^2 &=& e \left( \unitvec{A}{L}\cdot  {\mf u}_A - \nu\omega_D^* - (\unitvec{A}{R}\cdot  {\mf u}_A+ \nu\omega_D^*)\right)\\
&=& e \left(\left(\unitvec{A}{L} - \unitvec{A}{R} \right)\cdot  {\mf u}_A - 2\nu\omega_D^*\right)
\eqn
Recalling that $\unitvec{A}{L}$ and $\unitvec{A}{R}$ are unit vectors, notice that $\| \unitvec{A}{L} - \unitvec{A}{R} \|\leq 2$, and the equality holds when $\unitvec{A}{L} = -\unitvec{A}{R}$, which can be true only when $\xa$ is on the perimeter.
Therefore, we have the bound $|\left(\unitvec{A}{L} - \unitvec{A}{R} \right)\cdot  {\mf u}_A|<2\nu$, which gives
\bqn
\frac{\nu}{2}\frac{d}{dt}e^2 &=& |e|\text{sgn}(e) \left( \left(\unitvec{A}{L} - \unitvec{A}{R} \right)\cdot  {\mf u}_A-2\nu\text{sgn}(e)\right)\\
&<& -|e|\left( -2\nu\text{sgn}(e) + 2\nu\right)\\
&\leq& 0.
\eqn
Therefore, the error is stabilized around 0, implying that $J_L^*=J_R^*$, i.e., $\xa\in\affsurface(s_D)$.
$\blacksquare$
\end{pf}

Since the afferent surface extends from the defender's position, the lemma shows that the intruder can only reach the perimeter by passing through the defender position: i.e., it cannot reach the perimeter without getting captured.
Therefore, we extend the definition of capture from $\xa = \curve(s_D)$ to the condition $\xa\in\affsurface(s_D)$, and use it as part of the terminal condition.
Note that the former condition is contained in the latter.

\begin{lemma}\label{lem:defender_winning_region}
Let $\dwinr(s_D)$ denote the complement of $\awinr(s_D)$.
If the initial condition is $\xa\in\dwinr(s_D)$, i.e., $\mf x_A\notin\awinr(s_D)$, then regardless of the intruder strategy, the defender wins the game of kind using $\omega_D^*$ in \eqref{eq:omega_star}: i.e., the defender either captures the intruder or prevents it from scoring indefinitely.
\end{lemma}
\begin{pf}
Suppose the intruder never enters the winning region $\awinr$.
Then, since $\awinr$ contains the entire perimeter other than a single point $s_D$ (defender position), the only entry point to the perimeter is now $s_D$.
However, entering the perimeter from $s_D$ means capture.
Therefore, for the intruder to win the game, it is necessary to enter $\awinr$.
The question is: can the intruder start outside of $\awinr$ and enter it?

Crossing the boundary $\partial \awinr$ and entering $\awinr$ requires $V(s_D,\xa)$ to increase from negative to positive.
However, this is impossible when $\xa\in\Omega_L(s_D)$ because
\bnl
\dot{V} = \dot{J}_L^* = \frac{1}{\nu}\unitvec{A}{L}\cdot {\mf u}_A - \omega_D^* \leq  0.
\enl
We similarly have $\dot{V}\leq 0$ for $\xa \in \Omega_R(s_D)$. 
Therefore, $V(s_D,\xa)$ is non increasing, and so the intruder cannot enter the region $V>V(t_0)$, implying that it cannot enter $\awinr$.
$\blacksquare$
\end{pf}

The results of this section is summarized in the following theorem:
\begin{theorem}
The zero level set of $V(s_D,\xa)$ defined in \eqref{eq:value} gives the barrier of the game of kind.
\end{theorem}
The result directly follows from Lemmas~\ref{lem:intruder_winning_region}, \ref{lem:stability_around_affsurface} and \ref{lem:defender_winning_region}.
\if\ARXIVversion0
{\color{\maincolor} The algorithmic forms of the strategies are provided in the unabridged version \cite{shape_long}.}
\fi

\if\ARXIVversion1
{\color{\arxivcolor}
We also provide the intruder and defender strategies in the algorithm form.
The key step for both strategies is to determine whether the intruder is in the left region $\Omega_L$ or in the right region $\Omega_R$.
Importantly, this question can be answered without explicitly calculating the boundaries of the regions:}
 \begin{algorithm}[h]
{\color{\arxivcolor}
 \caption{Determining region (1 vs.\ 1) \label{alg:left}}
 \begin{small}
 \begin{algorithmic}[1]
\State {\bf Input}: $s_D$, $\xa$, $\curve$, and $\nu$
\State Compute $s_L$ and $s_R$ using Alg.~\ref{alg:sL}
\State $J_L^*\gets J_L(s_L;s_D,\xa)$ using \eqref{eq:JL}
\State $J_R^*\gets J_R(s_R;s_D,\xa)$ using \eqref{eq:JR}
\If{ any of the conditions in \eqref{eq:regions1v1} is true}
    \State{\texttt{is\_in\_Left} $\gets True$}
\Else
    \State{\texttt{is\_in\_Left} $\gets False$}
\EndIf
\State {\bf Return}: \texttt{is\_in\_Left}
\end{algorithmic}
\end{small}
}
\end{algorithm}

{\color{\arxivcolor}
Given the information \texttt{is\_in\_Left}, we can immediately calculate the control input:
}
 \begin{algorithm}[h]
{\color{\arxivcolor}
 \caption{Intruder control (1 vs.\ 1) \label{alg:intruder1v1}}
 \begin{small}
 \begin{algorithmic}[1]
\State {\bf Input}: $s_D$, $\xa$, $\curve$, and $\nu$
\State Compute $s_L$ and $s_R$ using Alg.~\ref{alg:sL}
\State Determine the region (i.e., \texttt{is\_in\_left}) using Alg.~\ref{alg:left} 
\If{ \texttt{is\_in\_left} $=True$}
    \State{$\mf u_A^* \gets \nu \hat{\mf{x}}_{A/L}$}
\Else
    \State{$\mf u_A^* \gets \nu \hat{\mf{x}}_{A/R}$}
\EndIf
\State {\bf Return}: $\mf u_A^*$
\end{algorithmic}
\end{small}
}
\end{algorithm}

 \begin{algorithm}[h]
{\color{\arxivcolor}
 \caption{Defender control (1 vs.\ 1) \label{alg:defender1v1}}
 \begin{small}
 \begin{algorithmic}[1]
\State {\bf Input}: $s_D$, $\xa$, $\curve$, and $\nu$
\State Determine the region (i.e., \texttt{is\_in\_left}) using Alg.~\ref{alg:left} 
\If{ \texttt{is\_in\_left} $=True$}
    \State{$\omega_D^* \gets 1$}
\Else
    \State{$\omega_D^* \gets -1$}
\EndIf
\State {\bf Return}: $\omega_D^*$
\end{algorithmic}
\end{small}
}
\end{algorithm}
\fi

\subsection{Optimality of the Strategies} \label{sec:optimality1v1}
This section discusses how the strategy set $(\omega_D^*,\mf u_A^*)$ defined in \eqref{eq:omega_star} and \eqref{eq:ua_star} forms an equilibrium also in the \emph{game of degree} for some objective functions.
We visit intruder-winning and defender-winning configurations separately.

Suppose the initial configuration is $\xa \in \awinr(s_D)$.
Then consider the following objective function:
\bnl
\label{eq:objective1}
\objective_1(\omega_D,\mf u_A) = \min \{ \arclength{D}{B}(t_F),\arclength{B}{D}(t_F) \},
\enl
where $t_F$ is the time the intruder breaches the perimeter at point $s_B$.
This quantity $\objective_1$ describes the \emph{safe distance} at the time of breaching which the intruder maximizes and the defender minimizes.
The $\min$ operator is used to account for both ccw and cw measure of the distance.

\begin{theorem}\label{thm:optimality_a}
If the initial configuration satisfies $\mf x_A \in \awinr(s_D)$, and if the players use $\objectivea$ in \eqref{eq:objective1} as the objective function, then $\mf u_A^*$ in \eqref{eq:ua_star} and $\omega_D^*$ in \eqref{eq:omega_star} form an equilibrium, and the value of the game is $V(s_D,\xa)$ in \eqref{eq:value}:
\beq
V = \min_{\omega_D} \max_{\mf u_A} \objectivea(\omega_D,\mf u_A) = \max_{\mf u_A} \min_{\omega_D} \objectivea(\omega_D,\mf u_A).
\eeq
\end{theorem}
\vspace{-10 pt}
\begin{pf}
Suppose $\xa\in\Omega_L$ without the loss of generality.
Along the terminal surface $\{[s_D,\xa]\,|\,\xa\in\partial\target\}$, we have $\xa = \curve(s_B)$ where $s_B\in\sleft$ from the supposition.
We also have $V=J_L^*(s_D,\xa)=\arclength{D}{B}$ since the term $\|\curve(s_L)-\xa\|$ in \eqref{eq:JL} is 0.
Noting that $\arclength{D}{B} = \min\{\arclength{D}{B},\arclength{B}{D}\}$ for $s_B\in\sleft$, we have $\objectivea = J_L^*$ along the terminal surface.
Therefore, maximizing or minimizing $\objectivea$ is equivalent to maximizing or minimizing $J_L^*(t_F)$ on the terminal surface.
Recalling the time derivative in \eqref{eq:JLdot}, we have
\bqn
1 &=& \argmin_{\omega_D} \max_{\mf u_A} \dot{J}_L^*(\omega_D,\mf u_A)\\
\nu\unitvec{A}{L} &=& \argmax_{\mf u_A} \min_{\omega_D} \dot{J}_L^*(\omega_D,\mf u_A)
\eqn
\vspace{-5 pt}
and
\bnl
\min_{\omega_D} \max_{\mf u_A}\dot{J}_L^*(\omega_D,\mf u_A) &=&
\max_{\mf u_A} \min_{\omega_D}\dot{J}_L^*(\omega_D,\mf u_A) \notag\\
&=&\dot{J}_L^*(1,\nu \unitvec{A}{L}) = 0.
\enl
The above results prove the theorem.
$\blacksquare$
\end{pf}

\begin{remark}
A similar result will be obtained for any objective function that is an increasing function of $\objectivea$.
For example, let $\alpha:[0,L/2)\rightarrow[0,\infty)$ be a strictly increasing function.
Then $\objective' \triangleq \alpha(\objectivea)$ is a valid objective function that has $\mf u_A^*$ in \eqref{eq:ua_star} and $\omega_D^*$ in \eqref{eq:omega_star} as the equilibrium strategies.
The value of the game is then $V'=\alpha(V)$.
The proof relies on the fact that $\objectivea = V$ along the terminal surface and $\dot{V}=0$ everywhere under the optimal strategies.
\end{remark}

\begin{remark}
If the intruder's objective is to quickly reach the perimeter, e.g., $P'= -(t_F-t_0)$, then the optimal intrusion strategy will be different.
In this case, the intruder will move straight towards the closest point on the perimeter whenever it avoids capture.
Otherwise, it will choose the breaching point so that $\objectivea=\varepsilon$, instead of maximizing the safe distance.
\end{remark}

\begin{remark}
The shortest path towards any $s_B\notin\sdirect$ consists of a straight line towards the tangent point and the path along the perimeter, which is equivalent to breaching the perimeter at the tangent point.
Therefore, it is reasonable for the intruder to choose $s_B \in \sdirect$.
\end{remark}

In the defender winning scenario, we can consider the following quantity which describes the distance of the intruder from the barrier:
\bnl
\distbarrier = \min\limits_{\mf x\in \awinr(s_D)} \|\mf x - \mf x_A\|.
\enl
This quantity can be interpreted as a buffer / margin from the intruder winning configuration.
The defender will want to maximize this buffer, whereas the intruder can minimize $\distbarrier$ hoping that any ``mistake'' in defender's behavior will let it penetrate the barrier and enter $\awinr(s_D)$.

Let the terminal payoff function to be the negative of the distance from the barrier when the capture occurs at time $t_F$:
\bnl
\label{eq:objective2}
\objectiveb(\omega_D,\mf u_A) \triangleq -\distbarrier(t_F) < 0.
\enl
Note that capture is defined by $\xa\in\affsurface(s_D)$ (see the paragraph before Lemma~\ref{lem:defender_winning_region}).
The defender tries to minimize $\objectiveb$, while the intruder tries to maximize it.


\begin{theorem}\label{thm:optimality_d}
If the initial configuration is $\mf x_A \notin \awinr(s_D)$, and if the players use $\objective_2$ in \eqref{eq:objective2} as the objective function,  then $\mf u_A^*$ in \eqref{eq:ua_star} and $\omega_D^*$ in \eqref{eq:omega_star} form equilibrium strategies, and the value of the game is $V(s_D,\xa)$ in \eqref{eq:value}:
\beq
V = \min_{\omega_D} \max_{\mf u_A} \objectiveb(\omega_D,\mf u_A) = \max_{\mf u_A} \min_{\omega_D} \objectiveb(\omega_D,\mf u_A).
\eeq
\end{theorem}
\vspace{-10 pt}
\begin{pf}
Following the proof of Theorem~\ref{thm:optimality_a}, it is sufficient to show the following identity:
\bnl
-\distbarrier(s_D,\xa) = 
\left\{ \begin{array}{r l}
\nu J_L^*(s_D,\xa) & \text{\;\;if\;\;} \mf x_A \in \Omega_L(s_D)\\
\nu J_R^*(s_D,\xa)& \text{\;\;otherwise}.
\end{array}\right.
\enl
In the following we prove the case with $\xa \in \Omega_L(s_D)$.

\begin{figure}
\begin{center}
\includegraphics[width=.49\textwidth]
{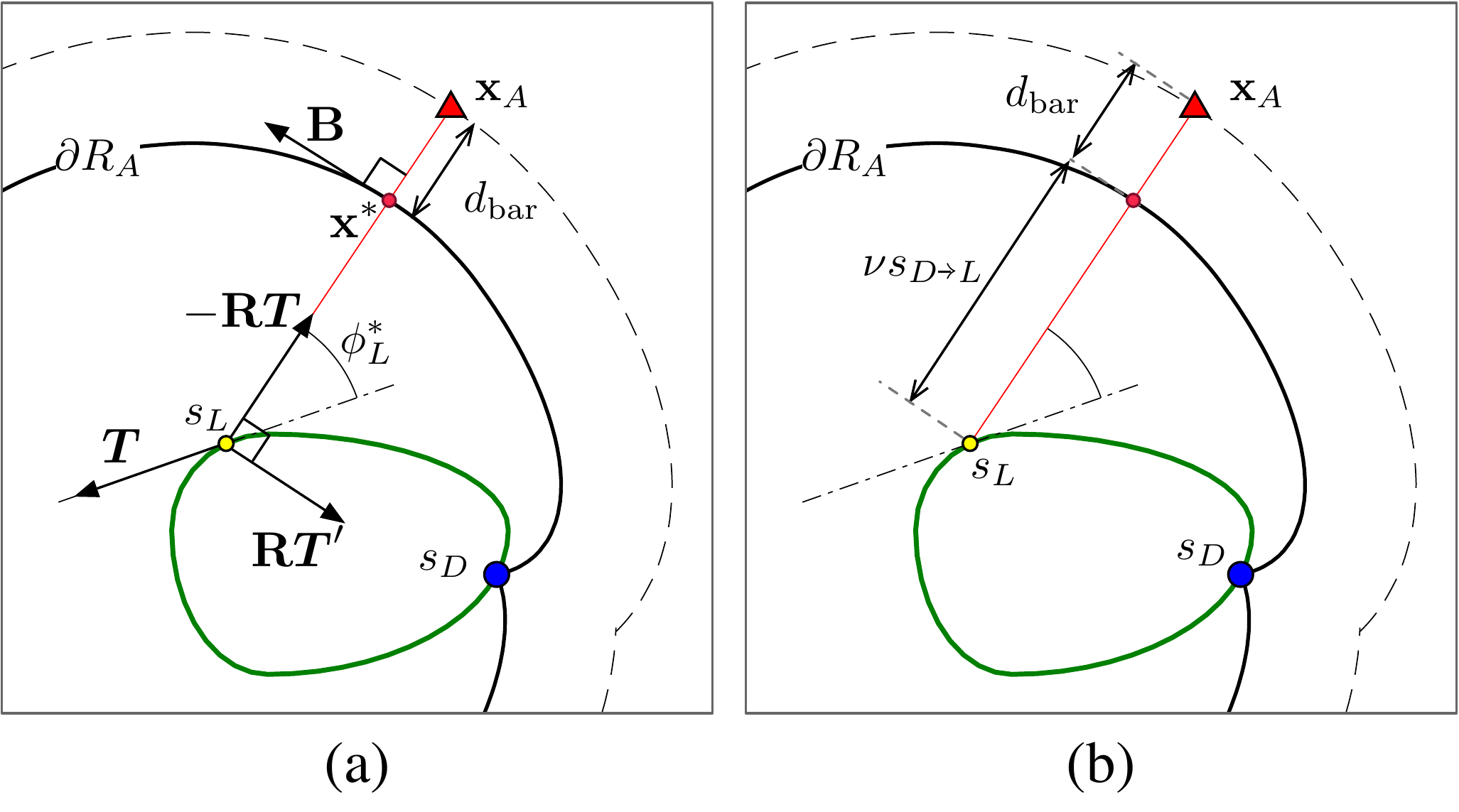}    
\caption{
Proof of the identity $ \distbarrier = -\nu J_L^*$.
}  
\label{fig:dbar}                                 
\end{center}                                 
\end{figure}
We first show that the point $\mf x^*$, which is the point in $\awinr$ that is closest from $\xa$, lies on the straight line from $\xa$ to $\curve(s_L)$ (see Fig.~\ref{fig:dbar}).
It suffices to show that $\relvec{A}{L}$ is perpendicular to the tangent of the barrier $\partial \awinr$ at $\mf x^*$, denoted by $\bf B$.

We can treat $s_L$ as a parameter to express a point, $\mf x_\text{bar}$, on the left barrier $\partial \awinr$ as follows:
\bqn
\mf x_\text{bar}(s_L) = \curve(s_L) - \nu \arclength{D}{L} \rotmat\tangent(s_L),
\eqn
where $\rotmat\in\mathbb{R}^{2\times 2}$ denotes the matrix for ccw rotation by $\phi^*_L$.
The tangent is obtained by
\bqn
\mf B \triangleq \frac{d\mf x_\text{bar}}{d s_L} = \tangent - \nu\rotmat\tangent- \nu\arclength{D}{L}\rotmat \tangent',
\eqn
where $\tangent'=\frac{d \tangent(s_L)}{d s_L}$ denotes the normal vector of $\curve$.
The inner product with $\unitvec{A}{L}$ gives
\bqn
\mf B\cdot \unitvec{A}{L} &=& \tangent\cdot\unitvec{A}{L} - \nu\rotmat\tangent\cdot\unitvec{A}{L} - \nu\arclength{D}{L}\rotmat \tangent' \cdot \unitvec{A}{L}\\
&=& \cos(\phi_L^*) - \nu (\unitvec{A}{L} \cdot \unitvec{A}{L}) - 0\\
&=& 0,
\eqn
where from the first to second line we used $\rotmat\tangent=\unitvec{A}{L}$ and $\rotmat \tangent'\cdot \unitvec{A}{L} =0$.

Now, the distance between $\xa$ and $\mf x^*$ is
\bqn
\distbarrier = \|\relvec{A}{L}\| -\nu\arclength{D}{L} = -\nu J_L^*.
\eqn
The case with $\xa\in\Omega_R$ can be shown similarly.
$\blacksquare$
\end{pf}

\vspace{-5 pt}
Unlike the intruder strategy, it is easy to see that the defender strategy will stay the same even if the objective is chosen to be the minimum time capture.

\subsection{Special Cases}
\label{sec:special_case}
This section discusses how the results provided in the preceding sections accommodate the two special cases considered in \cite{shishika2018cdc}: circular perimeter, and equal speed.

When the perimeter is a circle with radius $R$, the symmetry allows us to reduce the state space to $[r,\theta]$, where $r$ is the intruder's radial distance from the perimeter, and $\theta\in[-\pi,\pi]$ is the relative polar angle between the defender and the intruder with respect to the center of the circle.

\begin{figure}
\begin{center}
\includegraphics[width=.48\textwidth]
{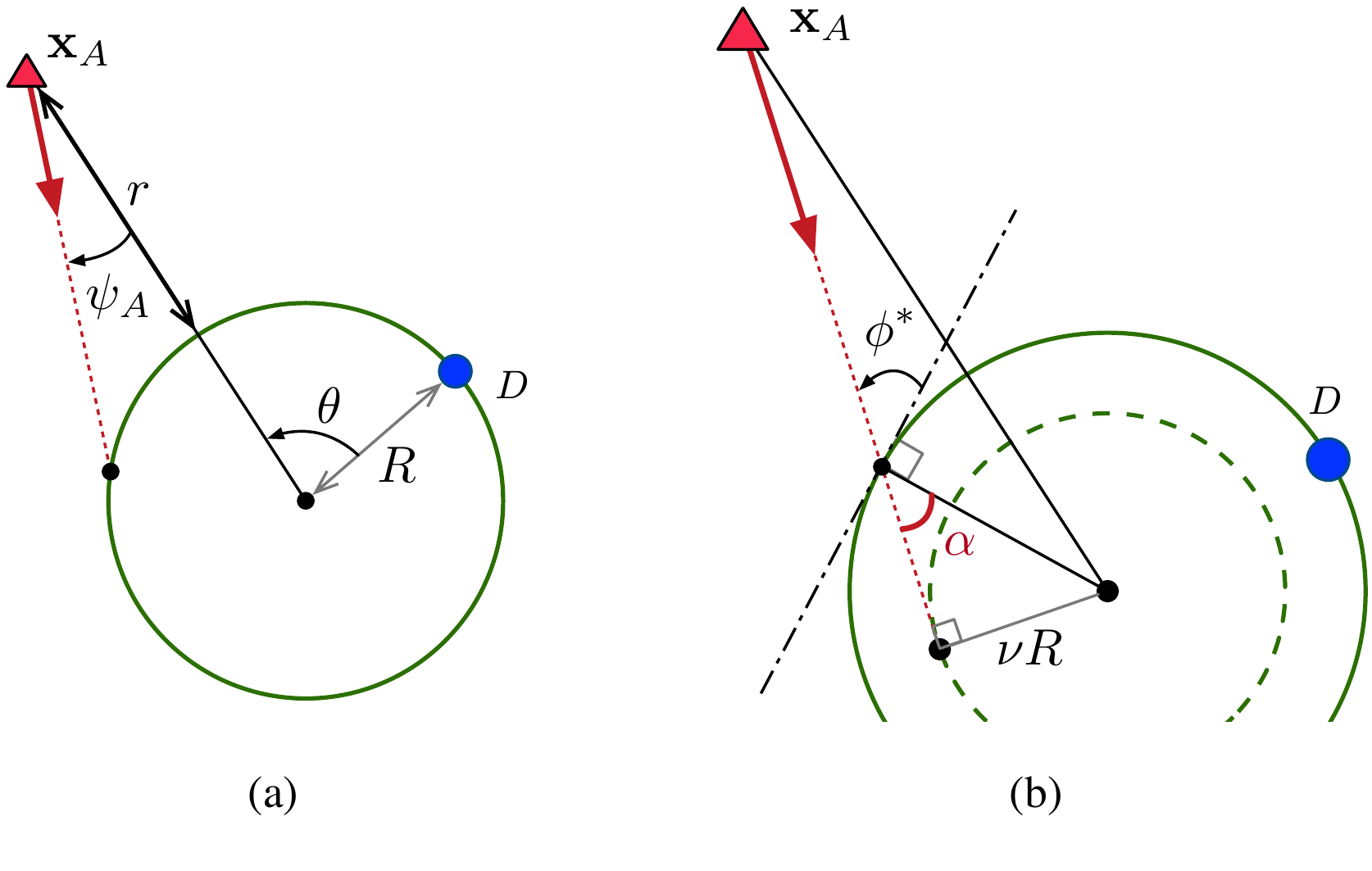}    
\caption{
Circular perimeter case. 
(a) States $[r,\theta]$ and the intruder's heading angle $\psi_A$.
(b) Computation of the approach angle $\phi^*$.
}  
\label{fig:circle}                                 
\end{center}                                 
\end{figure}

Whether the intruder is in the left region or the right region is determined by the sign of $\theta$: $\xa\in\Omega_L(s_D)$ if $\theta>0$, and $\xa\in\Omega_R(s_D)$ if $\theta<0$.
The singular surfaces correspond to the lines $\theta=0$ and $\theta=\pm\pi$.
The intruder control is parameterized by its speed $v_A$ and the heading $\psi_A$ as shown in Fig.~\ref{fig:circle}a.

\begin{theorem}[from \cite{shishika2018cdc}]
For a circular perimeter, the optimal strategies are
\emph{
\bnl
\omega_{D}^* &=& \text{sgn}(\theta),\;\text{and} \label{eq:d_strategy} \\
(v_A^*, \psi_A^*)  &=& \lp \nu,\,  \text{sgn}(\theta) \sin^{-1}\lp \frac{\nu R}{R+r}\rp \rp,\label{eq:phistar}
\enl
}
and the value of the game is
\emph{
\bnl
 V(r,\theta;\nu) = |\theta| - F(r)+F(0),
\label{eq:value_circ}
\enl
}
where
\bnl
F(r) = \sqrt{\lp\frac{R+r}{\nu R}\rp^2 - 1} - \cos^{-1}\lp \frac{\nu R}{R+r}\rp.
\enl
\end{theorem}
The sign function accommodates the switching between the left and the right regions.

The intrusion strategy allows further geometric interpretation: the optimal path of the intruder is to move towards the tangent point of the circle with radius $\nu R$ (see Fig.~\ref{fig:circle}b) \cite{shishika2018cdc}.
To verify this result with the strategy given in \eqref{eq:ua_star}, we compute the approach angle as follows.
The angle $\alpha$ in Fig.~\ref{fig:circle}b is $\alpha = \sin^{-1} \lp \frac{\nu R}{R} \rp = \sin^{-1} (\nu)$.
The approach angle is $\phi^* = \pi - \frac{\pi}{2}-\alpha = \frac{\pi}{2}-\sin^{-1} (\nu)$, which gives the relation $\phi^* = \cos^{-1}(\nu)$.
Recalling the results in \eqref{eq:partial_JL}, the circular case matches with our analysis in this paper.


The other special case is when the speed ratio is $\nu=1$.
Notice that the objective function now has the form $J_L^* = \arclength{D}{L} - \|\relvec{A}{L}\|$, in which case the level set $V=0$ is generated by the locus of intruder positions where $\|\relvec{A}{L}\| = \arclength{D}{L}$ (and similarly for the right breaching points).
In addition, recalling Remark~\ref{rem:extreme_cases}, the optimal breaching points are $s_L = \stanl $ and $s_R = \stanr$.
These properties are sufficient to see that the barrier $\partial\awinr$ is given by a curve called the \emph{involute} --- a locus of the tip of a taut string unwound from the geometry.
The left and the right part of the barrier corresponds to unwinding the string in ccw and cw directions.


\section{Two vs. One
Game}\label{sec:threeplayergame}

\noindent The next building block is the game played between two defenders ($\di$, $\dj$) and one intruder.
The states of the system are now $[\sdi,\sdj,\xa]$.
We follow the same structure as the previous section and discuss both the game of kind and the game of degree.

\subsection{Geometries}

A naive extension of the one vs.\ one game will conclude that the intruder will win if it is in the winning region against both defenders $\di$ and $\dj$, i.e., if $\xa$ is in
\bql
\awinri \triangleq \{ \mf x \, | \, V(\sdi,\mf x)>0 \text{\;and\;} V(\sdj,\mf x)>0\}.
\label{eq:independent_awinr}
\eql
The subscript $_I$ is used to reflect the fact that the games against $\di$ and $\dj$ are \emph{independently} considered.
However, in reality, the optimal intrusion strategy and the winning regions cannot be obtained by treating $\di$ and $\dj$ separately, since the intruder must avoid both $\di$ and $\dj$ simultaneously.

\begin{figure}
\begin{center}
\includegraphics[width=.49\textwidth]
{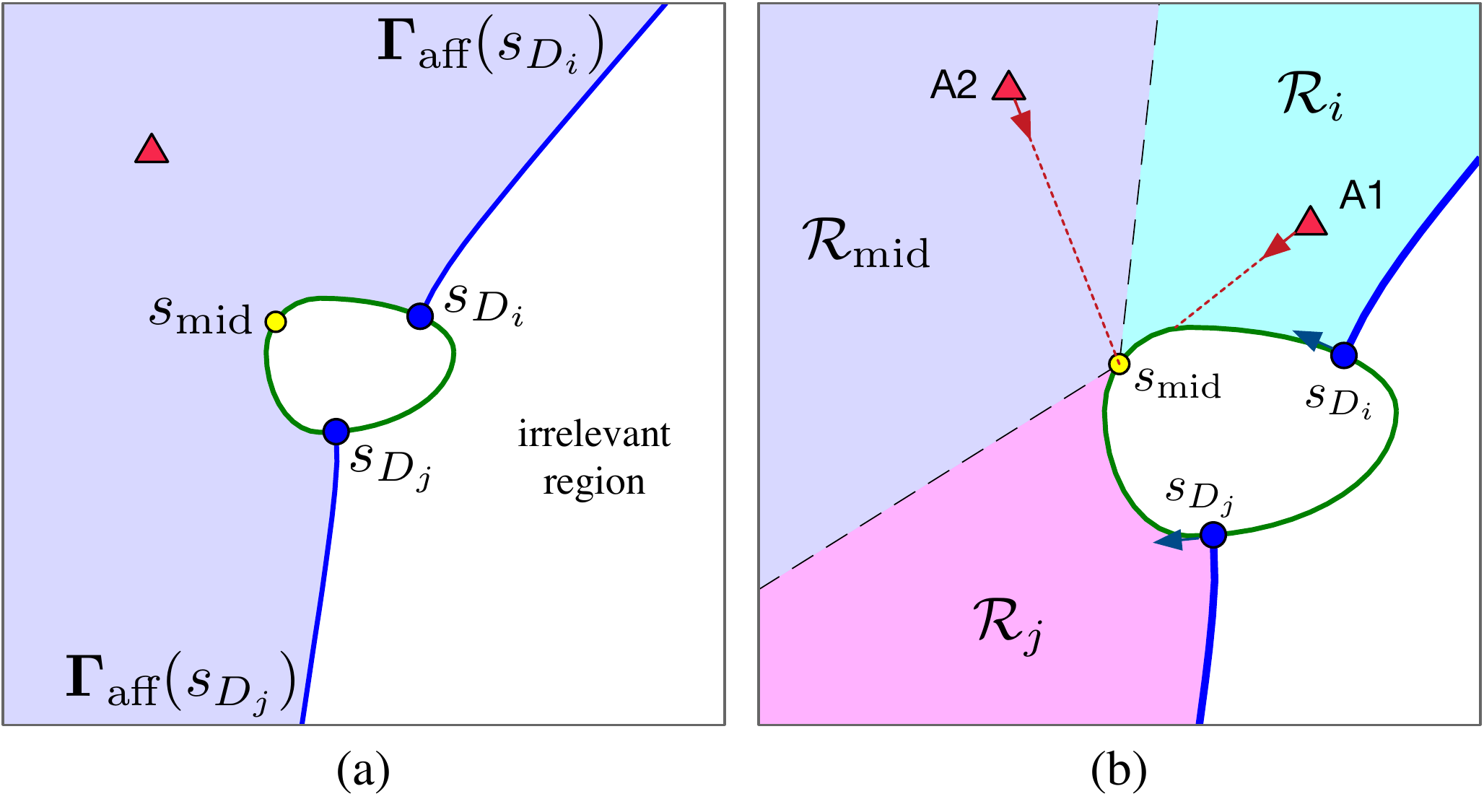}    
\caption{ 
Regions in the two vs.\ one game.
(a) Game space divided by the two afferent surfaces.
(b) Further division into three regions based on the location of the left and right breaching points.
}  
\label{fig:regions2v1}                                 
\end{center}                                 
\end{figure}

Observe that now the game space is divided into two parts by $\affsurface(\sdi)$ and $\affsurface(\sdj)$ (see Fig.~\ref{fig:regions2v1}a).
We showed in Sec.~\ref{sec:winning_regions} that the intruder cannot win if it reaches the afferent surface, so $\xa\in\affsurface(\sdi)\cup\affsurface(\sdj)$ is a part of the terminal condition.
Since $\xa$ cannot cross these surfaces, we focus our attention on the part of the game space that contains the intruder (shaded region in Fig.~\ref{fig:regions2v1}a) and ignore the other.
Without the loss of generality, we define $\di$ to be the one on the cw side and $\dj$ to be the one on ccw side (Fig.~\ref{fig:regions2v1}a).

The opposite point $\sDoppo$ was important in the one vs.\ one game because it was the farthest point from a single defender.
The analogy in the two vs.~one game is the midpoint, $\smid$, between the two defenders, which achieves the maximum distance from the nearest defender.

In deriving the intruder strategy, we consider the following quantity:
\bnl
\vspace{-5 pt}
J_{ij} = \min\{\arclength{\di}{B},\arclength{B}{\dj}\} - \frac{1}{\nu}\| \curve(s_B)-\xa  \|,
\enl
where the subscript $ij$ denotes the indices of the defenders.
The interpretation is similar to $J_L$ and $J_R$ in Sec.~\ref{sec:twoplayergame}.
It is the expected safe distance assuming that (i) $\di$ moves ccw, (ii) $\dj$ moves cw, and (iii) the intruder moves on a straight line path towards some breaching point $s_B\in[\sdi,\sdj]$.

For this function, we can consider three cases depending on where $s_B$ lies in:
\bnl
J_{ij} = \left\{
\begin{array}{ll}
J_L(s_B;\sdi,\xa) & \text{\;if\;} s_B\in [{\sdi}{\smid})\\
J_R(s_B;\sdj,\xa) & \text{\;if\;} s_B\in ({\smid}{\sdj}]\\
J_\text{mid}(\sdi,\sdj,\xa) & \text{\;otherwise: i.e.,\;} s_B=\smid,
\end{array}
\right.
\label{eq:J2v1}
\enl
where
\bnl
J_\text{mid} \triangleq \frac{1}{2}\arclength{\di}{\dj} - \frac{1}{\nu}\| \curve(\smid)-\xa \|
\enl
describes how much longer it takes for the defenders to reach $\smid$ than it does for the intruder.
The division described above is possible because only $\di$'s position is active in the calculation of $J_{ij}$ when the breaching point is in $[{\sdi},{\smid})$, and similarly only $\dj$'s position matters when $s_B\in({\smid},{\sdj}]$.

Following the above decomposition, the game space that contains the intruder can be further divided into three regions (see Fig.~\ref{fig:regions2v1}b):
\bnl
\label{eq:regions2v1}
\mc{R}_i &=& \{\xa \,|\, s_L \in[{\sdi}{\smid}) \}, \notag\\
\mc{R}_j &=& \{\xa \,|\, s_R \in({\smid}{\sdj}] \}, \text{\;and} \\
\mc{R}_\text{mid} &=& \{\xa \,|\,s_L \notin[{\sdi}{\smid}), s_R \notin({\smid}{\sdj}] \}.\notag
\enl
If $\xa \in \mc{R}_i(\sdi,\sdj)$, the intruder can move towards $s_L$ to play optimally against $\di$ without considering $\dj$, since $\sdj$ will not be active in $J_{ij}$.
Similarly when $\xa \in \mc{R}_j(\sdi,\sdj)$, the intruder can ignore $\di$ and choose $s_R$ to play optimally against $\dj$.
However, when $\xa \in \mc{R}_\text{mid}$, the intruder cannot simply choose one defender to play against because the optimal behavior against $\di$ makes $\dj$ to be the active defender and vice versa.
A good compromise in this case is to approach $\smid$.

Now we have a candidate intrusion strategy:
\bnl
\acontrol^* = \nu \unitvec{A}{\text{opt}},
\label{eq:acontrol2v1}
\enl
where $\unitvec{A}{\text{opt}}=\frac{\curve(\sopt)-\xa}{\|\curve(\sopt)-\xa\|}$, and the optimal breaching point is defined by
\bnl
\sopt(\mf x_A,s_{D_1},s_{D_2}) = \left\{
\begin{array}{l l}
s_L & \text{if\;} \xa\in\mc{R}_{i}(\sdi,\sdj) \\
s_R & \text{if\;} \xa\in\mc{R}_{j}(\sdi,\sdj)\\
\smid & \text{otherwise.}
\end{array}
\right.
\label{eq:sopt2v1}
\enl
The associated value (to be proved in Theorem~\ref{thm:optimality2v1}) is given as follows:
\bnl
V_{ij} = \left\{
\begin{array}{l l}
J_L^*(\sdi,\xa) & \text{if\;} \xa\in\mc{R}_{i}(\sdi,\sdj)\\
J_R^*(\sdj,\xa) & \text{if\;} \xa\in\mc{R}_{j}(\sdi,\sdj)\\
J_\text{mid}(\sdi,\sdj,\xa)& \text{otherwise},
\end{array}
\right.
\label{eq:value2v1}
\enl
where the regions are defined in \eqref{eq:regions2v1}.
\begin{figure}
\begin{center}
\includegraphics[width=.49\textwidth]
{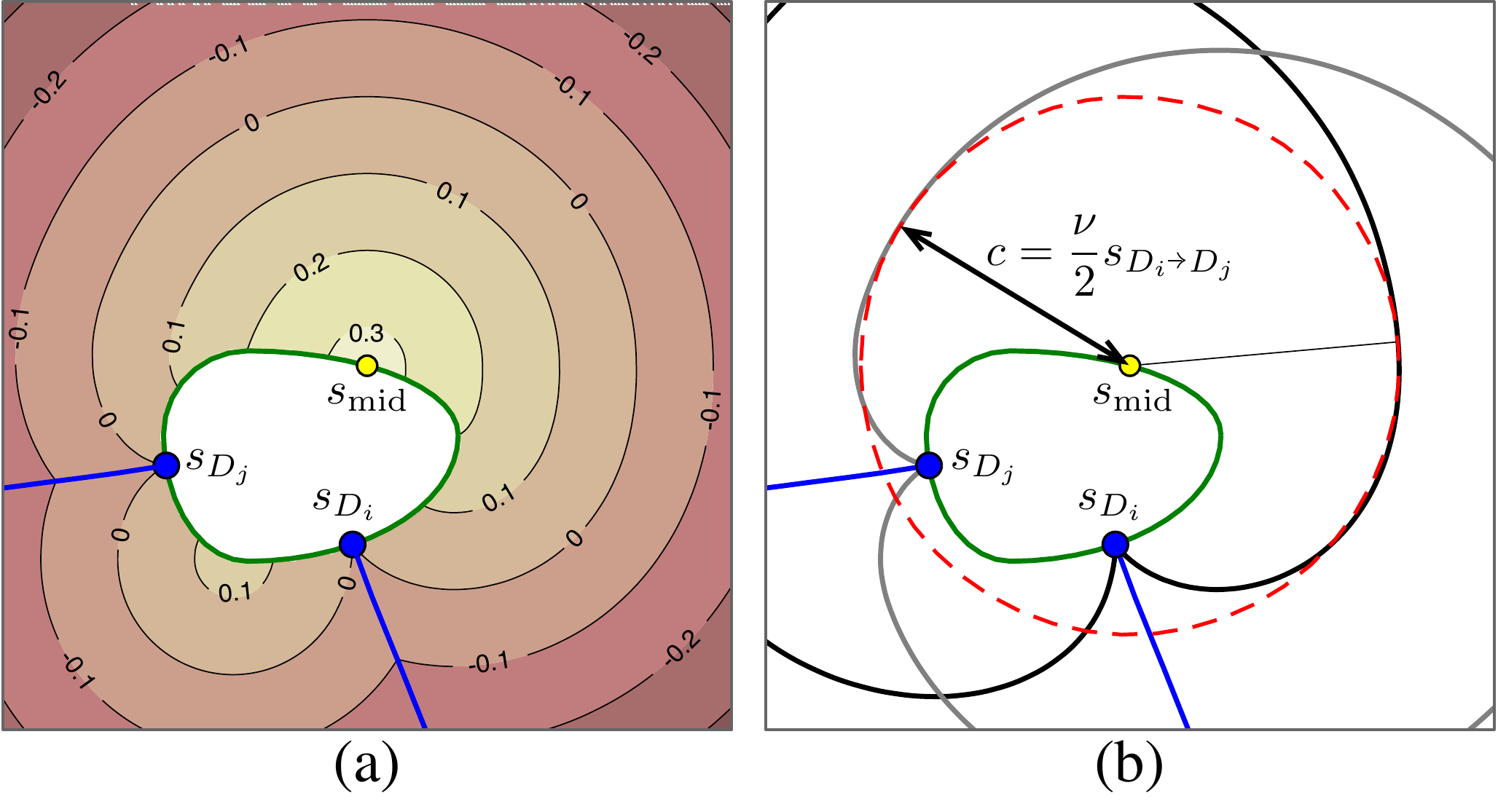}    
\caption{
(a) Level set of $V_{ij}$.
(b) Geometric construction of the zero level set of $V_{ij}$.
}  
\label{fig:levelset2v1}                                 
\end{center}                                 
\end{figure}
Fig.~\ref{fig:levelset2v1}a shows the level sets of $V_{ij}(\sdi,\sdj,\xa)$.
Each level set is a combination of three curves: the two level sets from the one vs.~one games and a circle centered at $\smid$.
Specifically, the zero level set $\{\xa\,|\,V_{ij}(\sdi,\sdj,\xa)=0\}$ is a combination of the two barriers, and a circle with radius
\bql
c=\frac{1}{2}\nu\arclength{\di}{\dj},
\eql
\vspace{-5 pt}
as shown in Fig.~\ref{fig:levelset2v1}b.

\subsection{Winning Regions}

Analogous to the two-player game, we define the intruder winning region to be the superlevel set of $V_{ij}$ as follows:
\bnl
\awinrc(\sdi,\sdj) \triangleq \{\xa\,|\,V_{ij}(\sdi,\sdj,\xa)>0  \}.
\enl
The subscript $_C$ is used to highlight the \emph{cooperative} nature of the associated defense strategy.
The following lemma gives a sufficient condition for intruder's victory:
\begin{lemma}\label{lem:intruder_winning_region2}
If the initial configuration satisfies $\xa(t_0)\in\awinrc(\sdi(t_0),\sdj(t_0))$, then regardless of the defender's strategy, the intruder wins the game of kind using $\acontrol^*$ defined in \eqref{eq:acontrol2v1}.
\end{lemma}
\vspace{-5 pt}
\begin{pf}
If the intruder starts in $\mc{R}_i\cap \awinrc$, then it wins against $\di$ by approaching $s_L$ since $J_L^*(\sdi,\xa)>0$.
In this case, although $s_L$ is suboptimal against $\dj$, the intruder still wins because $\arclength{L}{\dj}(t_0)>\arclength{\di}{L}(t_0)$: $\dj$ is farther from $s_L$ than $\di$.
With the same argument, the intruder wins if it starts in $\mc{R}_j\cap \awinrc$.
Finally, if $\xa \in \mc{R}_\text{mid}\cap \awinrc$, then the intruder can reach $\smid$ before either of the defenders because $J_\text{mid}>0$.
$\blacksquare$
\end{pf}

\vspace{-5 pt}
Observe that the intruder-winning region $\awinrc$ is smaller than $\awinri$ derived from the one vs.\ one game analysis.
The gap is generated by the cooperation between the defenders.
\begin{definition}
The \textbf{paired-defense region} is defined by:
\emph{
\bql
\dwinrp(\sdi,\sdj) = \awinri(\sdi,\sdj) - \awinrc(\sdi,\sdj).
\eql
}
\end{definition}
\begin{figure}
\begin{center}
\includegraphics[width=.49\textwidth]
{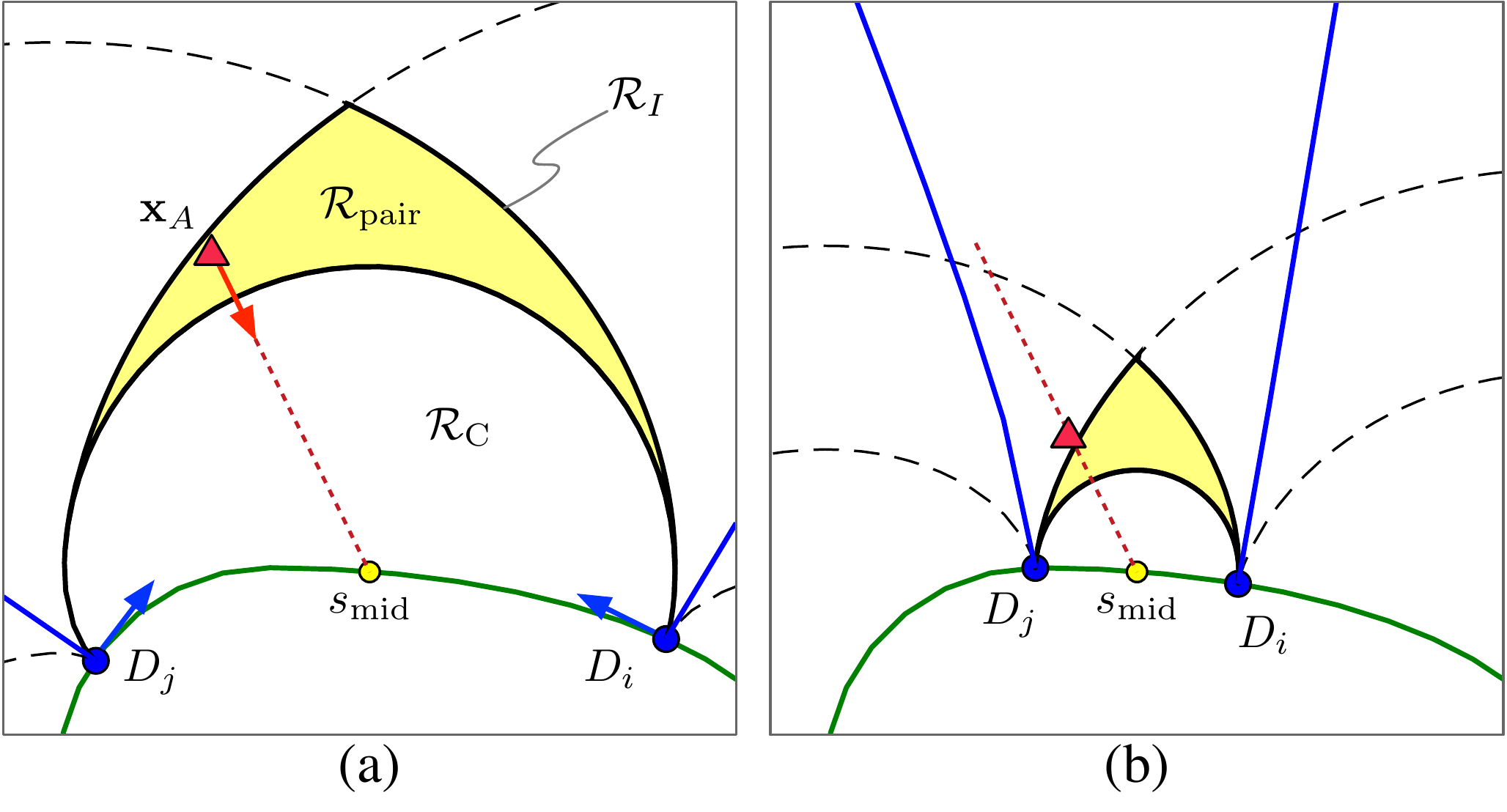}    
\caption{
Paired-defense region.
(a) Intruder starts in $\dwinrp$.
Neither $\di$ nor $\dj$ has a guarantee to win from the one vs.\ one game analysis because $\xa\in\awinri$.
(b) Pincer maneuver by the defender pair pushes the intruder out from $\dwinrp$, while also preventing it to enter $\awinrc$.
At this time, $\dj$ can guarantee its victory using one vs.\ one strategy since $\xa \in \dwinr(i)$.
}  
\label{fig:barrier2v1}                                 
\end{center}                                 
\end{figure}
\vspace{-5 pt}
\rev{The cooperation arises in the form of ``pincer movement,'' which is a tactic where the two defenders approach the intruder from both cw and ccw sides at the same time.
In our problem the corresponding control input is $[\omega_{\di},\omega_{\dj}]=[1,-1]$.}
By considering this defender strategy, the next lemma shows that $\xa\in\awinrc$ is also a necessary condition for the intruder to win the game of kind:

\begin{lemma}\label{lem:defender_winning_region2}
If the initial configuration satisfies $\xa\in\dwinrp(\sdi,\sdj)$, and if the defender pair uses a pincer movement, $[\omega_{\di},\omega_{\dj}]=[1,-1]$, then either $\xa\in\dwinr(\sdi)$ or $\xa\in\dwinr(\sdj)$ occurs before the intruder reaches the perimeter: i.e., the defender pair wins.
\end{lemma}
\vspace{-5 pt}
\begin{pf}
Observe that $\dwinrp$ shrinks as the two defenders get closer, and it disappears when the two meet at the midpoint.
Hence, the intruder will exit $\dwinrp$ in finite time.
There are only three ways to exit $\dwinrp$: enter $\dwinr(\sdi)$, enter $\dwinr(\sdj)$, or enter $\awinrc(\sdi,\sdj)$.
However, the intruder cannot enter $\awinrc$ because its speed $\nu$ cannot exceed the rate at which the radius of the circle decreases: $\dot{c}=\frac{1}{2}\nu\frac{d}{dt}\arclength{\di}{\dj} = \frac{1}{2}\nu(-1-1)=-\nu$.
Therefore, $\xa$ enters either $\dwinr(\sdi)$ or $\dwinr(\sdj)$.
$\blacksquare$
\end{pf}

\vspace{-5 pt}
Recalling that $\xa\in\dwinr(\sdi)\cup\dwinr(\sdj)$ trivially leads to capture based on the solution to the one vs.\ one game, the only region that the intruder can guarantee its victory is $\awinrc(\sdi,\sdj)$.

\begin{theorem}
The zero level set of $V_{ij}$ defined in \eqref{eq:value2v1}
gives the barrier of the game of kind played between two defenders and one intruder.
\end{theorem}
\vspace{-5 pt}
The result directly follows from Lemmas~\ref{lem:stability_around_affsurface}, \ref{lem:defender_winning_region2} and \ref{lem:intruder_winning_region2}.

\rev{
\vspace{-10 pt}
\begin{remark}
The two vs.\ one scenarios have also been studied in related but different problems.
Cooperative capture in pursuit-evasion games have been studied in \cite{Makkapati2018,Foley1974a,Garcia2017}.
There are also works that consider two vs. one cooperation in border-defense type scenarios \cite{garcia2019cooperative,yan2018twovsone}.
\end{remark}
}

\subsection{Optimality of the Strategies}
Consider the intruder winning configuration.
The payoff function in \eqref{eq:objective1} can be modified to
\bql
\label{eq:objective1b}
\objective_1(\omega_{\di},\omega_{\dj},\mf u_A) = \min \{ \arclength{\di}{B}(t_F),\arclength{B}{\dj}(t_F) \},
\eql
which describes the safe distance at the time of breaching.
\begin{theorem}\label{thm:optimality2v1}
If the initial configuration is $\xa\in\awinrc(\sdi,\sdj)$, and if the players use $\objective_1$ in \eqref{eq:objective1b} as the objective function, then $\mf u_A^*$ in \eqref{eq:acontrol2v1}, \eqref{eq:sopt2v1} and the pincer maneuver $[\omega_{\di},\omega_{\dj}]=[1,-1]$ form equilibrium strategies, and the value of the game is $V_{ij}$ in \eqref{eq:value2v1}.
\end{theorem}
\begin{pf}
Similar to the proof of Theorem~\ref{thm:optimality_a}, we can see that $\objectivea = V_{ij}$ along the terminal surface.
Therefore, the increase (resp.~reduction) in $\objectivea$ is equivalent to the increase (resp.~reduction) in $V_{ij}(t_F)$.
To prove the optimality, we will show that 
\bnl
\dot{V}_{ij}(\bs\omega_D^*,\mf u_A) \leq \dot{V}_{ij}(\bs\omega_D^*,\mf u_A^*)=0 \leq \dot{V}_{ij}(\bs\omega_D,\mf u_A^*),
\label{eq:Vijdot}
\enl
where $\bs\omega_D = [\omega_{\di},\omega_{\dj}]$, and $\bs\omega_D^* = [1,-1]$.
The above inequality indicates that any unilateral change in the strategy will result in a suboptimal performance.

Recall that $V_{ij} = J_L^*$ when $\xa\in\mc{R}_i$.
In this case, the inequality is shown using the time derivative $\dot{J}_L^*$ in the proof of Theorem~\ref{thm:optimality_a}.
The case with $\xa\in\mc{R}_j$ is similarly straightforward.
However, the case $\xa\in\mc{R}_\text{mid}$ has not been considered yet.
For example, can the defenders move in the same direction $\bs\omega_D^* = [1,1]$ to move $\smid$ away from the intruder?
We will investigate this using the time derivative $\dot{V}_{ij}=\dot{J}_\text{mid}$:
\bqn
\dot{J}_\text{mid} &=& \frac{1}{2}(\omega_{\dj}-\omega_{\di}) - \frac{\unitvec{A}{\text{mid}}}{\nu}\cdot\left(\dot{s}_\text{mid}\tangent(\smid) - \mf u_A \right) \\
&=&\frac{1}{2}\left( (1-\beta)\omega_{\dj} - (1+\beta)\omega_{\di}\right) + \frac{1}{\nu}\unitvec{A}{\text{mid}}\cdot\mf u_A,
\eqn 
where we used $\dot{s}_\text{mid} = \frac{1}{2}(\omega_{\dj}+\omega_{\di})$ and defined
\bqn
\beta \triangleq \frac{\unitvec{A}{\text{mid}}\cdot\tangent(\smid)}{\nu} = \frac{\cos\phi(\smid)}{\nu}.
\eqn
From the conditions on $s_L$ and $s_R$ (see \eqref{eq:regions2v1}), the approach angle at $\smid$ satisfies $\phi_L^*\leq \phi(\smid)\leq \phi_R^*$ when $\xa\in\mc{R}_\text{mid}$.
Hence, we have $|\cos\phi(\smid)| <\nu$, or equivalently, $|\beta| < 1$ when $\xa\in\mc{R}_\text{mid}$.
Therefore, both $1-\beta$ and $1+\beta$ are positive, and we have
\bqn
[1,-1] &=& \argmin_{\bs\omega_D} \max_{\mf u_A} \dot{J}_\text{mid}^*(\bs\omega_D,\mf u_A),\\
\nu\unitvec{A}{\text{mid}} &=& \argmax_{\mf u_A} \min_{\bs\omega_D} \dot{J}_\text{mid}^*(\bs\omega_D,\mf u_A),
\eqn
and $\min_{\bs\omega_D} \max_{\mf u_A}\dot{J}_L^*(\bs\omega_D,\mf u_A)  = 0$,
which completes the proof.
$\blacksquare$
\end{pf}

For the defender winning configuration, we use the same payoff $\objectiveb$ in \eqref{eq:objective2}, with a modification on $\distbarrier$ as follows:
\bnl
\distbarrier = \min\limits_{\mf x\in \awinrc} \|\mf x - \mf x_A\|.
\enl

\begin{theorem}\label{thm:optimality2v1d}
If the initial configuration is $\xa\notin\awinrc(\sdi,\sdj)$, and if the players use $\objectiveb$ in \eqref{eq:objective2} as the objective function, then $\mf u_A^*$ in \eqref{eq:acontrol2v1} and the pincer maneuver $[\omega_{\di},\omega_{\dj}]=[1,-1]$ form equilibrium strategies, and the value of the game is $V_{ij}$ in \eqref{eq:value2v1}.
\end{theorem}
\begin{pf}
Similar to the proof of Theorem~\ref{thm:optimality_d}, it suffices to show that $-\distbarrier = \nu V_{ij}$, since we already have the result \eqref{eq:Vijdot}.
The identity for the case with $\xa\in\mc{R}_i$ or $\xa\in\mc{R}_j$ is already proved in Theorem~\ref{thm:optimality_d}.
When $\xa\in\mc{R}_\text{mid}$, it is easy to get the result $\mf B\cdot \relvec{A}{\text{mid}}=0$ recalling that the barrier $\partial \awinrc$ in this portion is a circle whose center is at $\smid$.
$\blacksquare$
\end{pf}

The optimal behavior of the defender at $\sdi$ against an intruder at $\xa$ may be different based on the existence of the third player $\sdj$.
In a one vs.\ one game $\di$ must decide between cw and ccw motion based on the location of $\xa$ with respect to the dispersal surface $\dispsurface(\sdi)$, and it is possible that the cw motion is optimal.
On the other hand, in a two vs.\ one game $\di$ (defined as the one on cw side) should always move ccw.

\if\ARXIVversion1
{\color{\arxivcolor}
As was done for the one vs.\ one game, we provide algorithms to obtain key quantities necessary to compute the strategies.
First, recall that a defender pair divides the game space into two parts (Fig.~\ref{fig:regions2v1}a). 
Given a pair of defenders $s_{D1}$ and $s_{D2}$, we must first determine which acts as the cw-side defender ($D_i$) and which acts as the ccw-side defender ($D_j$).
 \begin{algorithm}[h]
 \caption{Relevant region (2 vs. 1) \label{alg:relevant_region}}
 \begin{small}
 \begin{algorithmic}[1]
\State {\bf Input}: $s_{D1}$, $s_{D2}$, $\xa$, $\curve$, and $\nu$
\State Compute \texttt{is\_in\_left1} with $s_{D1}$ using Alg.~\ref{alg:left} 
\State Compute \texttt{is\_in\_left2} with $s_{D2}$ using Alg.~\ref{alg:left} 
\If{ $\arclength{D1}{D2} < \frac{L}{2}$ }
    \State{\texttt{is\_in\_D1D2} $\gets$ \texttt{is\_in\_left1} $and$ $\sim$\texttt{is\_in\_left2}}
\Else
    \State{\texttt{is\_in\_D1D2} $\gets$ \texttt{is\_in\_left1} $or$ $\sim$\texttt{is\_in\_left2}}
\EndIf
\State {\bf Return}: \texttt{is\_in\_D1D2}
\end{algorithmic}
Note: $\sim$ is a negation operator
\end{small}
\end{algorithm}

If \texttt{is\_in\_D1D2}$=True$, then the defender at $s_{D1}$ takes the role of $D_i$ as described in this section, but it will act as $D_j$ otherwise.
}

{\color{\arxivcolor}
 \begin{algorithm}[h]
 \caption{Intruder Control (2 vs. 1) \label{alg:intruder2v1}}
 \begin{small}
 \begin{algorithmic}[1]
\State {\bf Input}: $s_{D1}$, $s_{D2}$, $\xa$, $\curve$, and $\nu$
\State Compute $s_L$ and $s_R$ using Alg.~\ref{alg:left}
\State Compute \texttt{is\_in\_D1D2} using Alg.~\ref{alg:relevant_region}
\If{ \texttt{is\_in\_D1D2} $=True$}
    \State{$\sdi \gets s_{D1}$ and $\sdj \gets s_{D2}$}
\Else
    \State{$\sdi \gets s_{D2}$ and $\sdj \gets s_{D1}$}
\EndIf
\State Determine the region using \eqref{eq:regions2v1}
\If{ $\xa \in \mc{R}_i$ }
    \State{$\mf u_A^* = \nu \unitvec{A}{L}$}
\ElsIf{ $\xa \in \mc{R}_j$ }
    \State{$\mf u_A^* = \nu \unitvec{A}{R}$}
\Else
    \State{$\mf u_A^* = \nu \unitvec{A}{\text{mid}}$}
\EndIf
\State {\bf Return}: $\mf u_A^*$
\end{algorithmic}
\end{small}
\end{algorithm}
}

\revrev{The defender strategy is presented in Alg.~\ref{alg:defender2v1}.
If one defender can guarantee capture, then the behavior of the other defender is inconsequential. Therefore, we assign no action to that defender in this paper.
If a single defender cannot guarantee capture, i.e., $\xa\notin\dwinr(\sdi)\cup\dwinr(\sdj)$, then the defenders perform pincer movement.}

\revrev{Finally, note that the attacker's current region identified with \eqref{eq:regions2v1}, together with the quantities $s_L, s_R$ and $\smid$ are sufficient to find $V_{ij}$ in \eqref{eq:value2v1}}

{\color{\arxivcolor}
 \begin{algorithm}[h]
 \caption{Defender Control (2 vs.~1) \label{alg:defender2v1}}
 \begin{small}
 \begin{algorithmic}[1]
\State {\bf Input}: $s_{D1}$, $s_{D2}$, $\xa$, $\curve$, and $\nu$
 \If{$\xa\in\dwinr(s_{D1})$}
    \State $\omega_{D_1}\gets$ strategy from Alg.~\ref{alg:defender1v1}
    \State $\omega_{D_2}\gets$ 0 
 \ElsIf{$\xa\in\dwinr(s_{D2})$}
    \State $\omega_{D_1}\gets$ 0 
    \State $\omega_{D_2}\gets$ strategy from Alg.~\ref{alg:defender1v1}
 \Else
\State Compute $s_L$ and $s_R$ using Alg.~\ref{alg:left}
\State Compute \texttt{is\_in\_D1D2} using Alg.~\ref{alg:relevant_region}
\If{ \texttt{is\_in\_D1D2} $=True$}
    \State $[\omega_{D_1},\omega_{D_2}] \gets [1, -1]$
\Else
    \State $[\omega_{D_1},\omega_{D_2}] \gets [-1, 1]$
\EndIf
 \EndIf

\State {\bf Return}: $[\omega_{D_1},\omega_{D_2}]$
\end{algorithmic}
\end{small}
\end{algorithm}
}

\fi

\begin{remark}[Computation]
Importantly, the calculation of the optimal strategies and the value (for both one vs.\ one and two vs.\ one) do not require any explicit computation of the surfaces nor the regions.
A numerical search is performed only in the first step when finding the breaching points, which is also simple due to the monotonicity of the approach angle $\phi(s)$.
\rev{Also note that numerical methods proposed in \cite{Chen2014a,chen2017} requires us to solve the HJI PDE offline and store the solution (i.e., control inputs corresponding to all possible states), so the players can use this `lookup table' in the run time. Our method requires less memory because control inputs are computed online.}
\end{remark}

\if\ARXIVversion0
{\color{\maincolor} The algorithms to obtain key quantities necessary to compute the strategies are presented in the unabridged version \cite{shape_long}.
}
\fi

\if\ARXIVversion1
\section{Multiplayer Game} \label{sec:team_strategy}
This section discusses assignment-based defense policies when there are multiple players on both teams. 
\rev{These multi-agent policies rely on the barriers or the winning regions derived in the previous sections.}

We first review the assignment method (MM defense) proposed by Chen et al.~\cite{Chen2014a,Chen2014b} that only considers one vs.~one defense.
\rev{We then propose an extension (MIS defense) that directly incorporates cooperative two vs.\ one defense, which was first introduced in the conference version of this paper \cite{shishika2018cdc}.
We also briefly introduce a cooperative defense strategy (LGR defense) that has the strongest theoretical guarantees, which is presented in our separate publication \cite{shishika2020ral}.
Finally, we provide a discussion on the strengths and weaknesses of each policy.
}

\rev{The bounds $Q_{MM}$, $Q_{MIS}$ and $Q_{LG}$ that will be introduced in this section provide solution to Prob.~2 posed in Sec.~\ref{sec:problem_formulation}.
}

\subsection{Maximum Matching (MM) defense}
For a given initial configuration $\{ \mf x_{A_i} \}_{i=1}^{N_A}$ and $\{ s_{D_j} \}_{j=1}^{N_D}$, the defender-winning regions can be used to determine a set of intruders that each defender can win against: $D_j$ can be assigned to $A_i$ if $\mf x_{A_i}\notin \awinr(\sdj)$, or equivalently, $\xa\in\dwinr(\sdj)$.
Again, the defender wins by either capturing the intruder or delaying its intrusion indefinitely (see Sec.~\ref{sec:winning_regions}).

One can generate a bipartite graph with intruders and defenders as two sets of nodes.
Edges will be drawn from each defender to all the intruders that it can capture.
Matching in graph theory refers to finding a set of edges with no shared nodes.
Here, this restriction corresponds to the assumption that $D_j$ can only play an optimal two-player game against at most one intruder at a time.
Maximum-cardinality matching (MM) algorithms (see references in \cite{Chen2014a}) give such an edge set with maximum cardinality.

The edge set is used to assign at most one unique defender to each intruder.
If $\dj$ is assigned to $A_i$, then $\dj$ selects its strategy to be optimal against $A_i$.
The cardinality of the edge set, $N_\text{MM}^{cap}$, tells us that at least $N_\text{MM}^{cap}$ intruders will be captured.
The upper bound on the intruder score is then given by 
\bql
Q \leq Q_\text{MM}= N_A - N_\text{MM}^{cap}.
\eql
This method assumes that all defenders play independent games and ignores any cooperation with the teammates.

\subsection{Maximum Independent Set (MIS) defense}
Now we allow a defender pair to be assigned to a single intruder. Let $D_{(i,j)}$ denote a pair $(D_i, D_j)$.
\begin{figure}[t]
\centering
\includegraphics[width=.5\textwidth]
{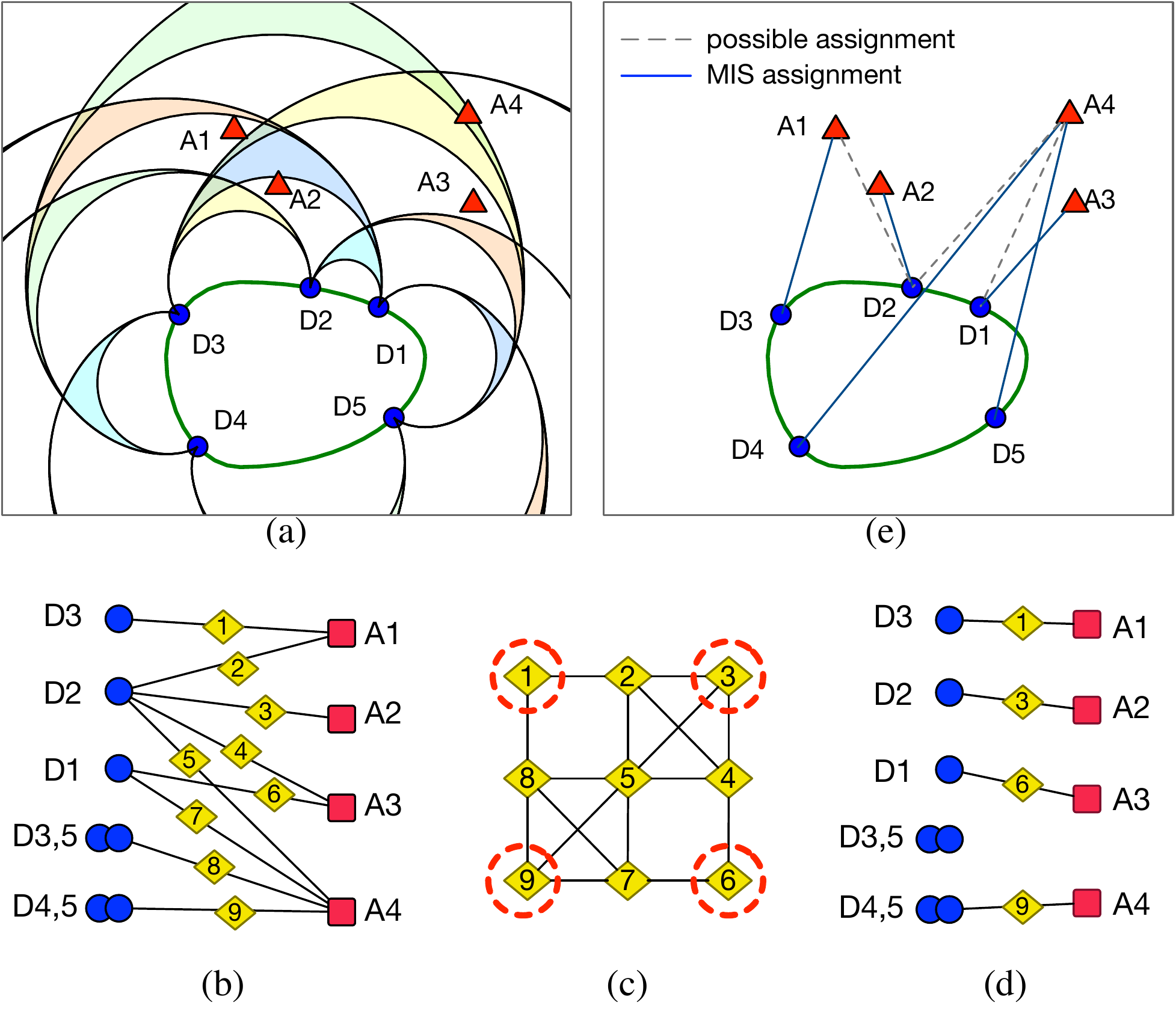}
\caption{(a) Example with 5 defenders and 4 intruders. (b) Each node on the left represents a defender or a pair of defenders, and nodes on the right represent intruders.  Edges are drawn when the defender or defender pair can win against the intruder. (c) Edges in (b) become nodes in the new graph.  A maximum independent set is highlighted in red.  (d) An assignment (not necessarily unique) that defends against maximum number of intruders. (e) Assignment described in the original game space.}
\label{fig:MIS}
\end{figure}
The matching algorithm needs to be modified to avoid conflicts.
For example, $D_i$ and a pair $D_{(i,j)}$ cannot be treated as independent nodes and be assigned to distinct intruders, because $D_i$ may not be able to move optimally against two intruders simultaneously.  
 We pose the assignment problem into a maximum independent set (MIS) problem \cite{KleinbergAlgorithm} as described in the following:
\begin{itemize}
\item[1)] Construct a bipartite graph with two sets of nodes $\mc{V}_D = \{{D_i} \}_{i=1}^{N_D}\cup\{D_{(i,j)} \}_{i\neq j} $ and $\mc{V}_A =\{A_i \}_{i=1}^{N_A}$.
The node set $\mc{V}_D$ now includes all possible defender pairs.
\item[2)] For each $D_i$, draw edges to all intruders, $A_k$, such that $\mf x_{A_k}\in\dwinr(\sdi)$.
\item[3)] For each pair $D_{(i,j)}$, draw edges to all $A_k$ such that $\mf x_{A_k}\in \dwinrp(\sdi,\sdj)$ (see Fig.~\ref{fig:barrier2v1}).
Note that we exclude the intruders that are independently capturable by either $D_i$ or $D_j$.
\end{itemize}
Figure~\ref{fig:MIS}a depicts a particular initial condition, and Fig.~\ref{fig:MIS}b shows the bipartite graph (nodes with no edges are omitted).  
\begin{itemize}
\item[4)] The edges in the graph are enumerated and become the nodes in the new graph representation  (see Fig.~\ref{fig:MIS}c).
\item[5)] Draw an edge between two nodes (in the new graph) whenever they share the same defender or intruder.
\item[6)] Find MIS, i.e., the largest subset of nodes with no direct connection.
\end{itemize}
Figures~\ref{fig:MIS}d-e illustrate the resultant assignments that give $N_\text{MIS}^{cap} = 4$ and $Q\leq\Qmis = N_A - N_\text{MIS}^{cap} = 0$.
Note that the maximum-matching assignment only guarantees $Q\leq \Qmm=1$ in this example.

Since the MIS formulation considers paired defense in addition to all the individual defenses, it gives equal or tighter upper bound for any initial configuration: i.e.,
\rev{
\begin{equation}
Q \leq Q_{MIS}\leq Q_{MM}.    
\label{eq:MIS_MM}
\end{equation}
The above result is also straightforward noting that bipartite matching problem can be encoded as a special case of the independent set problem \cite{KleinbergAlgorithm}.
}

The downside of the above formulation is the fact that MIS cannot be found efficiently  \cite{KleinbergAlgorithm}. 
\rev{While approximation methods to solve MIS exist (e.g., \cite{yan2019three}), they potentially make the inequality \eqref{eq:MIS_MM} to no longer hold, which takes away the whole purpose of using the MIS strategy.}
A computationally efficient team policy \rev{that preserves the effectiveness of the cooperative defense is presented next}.

{\color{black}
\subsection{Local Game Region (LGR) defense}
We finally present an approach that gives us the strongest theoretical guarantees.
The full detail of this policy is presented in our separate publication \cite{shishika2020ral}, and therefore we only provide a high-level idea here.

The core concept we use for this strategy is called the Local Game Region (LGR), which is defined by the intruder winning region in the two vs.\ one game, with an addition of a degenerate case where the two defenders are identical. There are $_{N_D}C_2 \times 2 = N_D(N_D-1)$ ordered pairs and $N_D$ degenerate cases resulting in $N_D^2$ regions in total. We use $k\in 1,...,N_D^2$ to denote the indices of the regions.

For each LGR, We can define an intruder and a defender subteams by collecting all intruders and defenders in the region. Let $n_A^k$ and $n_D^k$ denote the number of agents in the $k$th intruder and defender subteams.
Then we can define the \emph{numerical advantage} held by the intruder subteam as follows:
\begin{equation}
q_k = \max\{n_A^k-n_D^k, 0\}.    
\end{equation}

We call this number the \emph{local game score}.
The significance of this quantity is that we can prove that the intruder subteam can guarantee to score at least $q_k$ points by approaching near the mid point between the defender pair that defines this $k$th LGR \cite{shishika2020ral}.

Now considering the overall game, the intruders can maximize their score by selecting the optimal decomposition into subteams, i.e., a selection of a set of LGRs.
We show in \cite{shishika2020ral} that this team selection can be cast as the following optimization problem:
\begin{equation}
Q_{LG} = \max_{\bf G} \sum_{k\in \bf G} q_k,
\label{eq:QLG}
\end{equation}
where $\bf G$ denotes a set of disjoint LGRs, that does not share any  area.

Conveniently, the optimal disjoint set $\bf G^*$ and the value $Q_{LG}$ can be obtained in $O(N_D^4)$ time by recognizing \eqref{eq:QLG} as an instance of the maximum weight independent set problem on a circular arc graph \cite{shishika2018cdc}.
For applications where it is critical to avoid any intrusion, it is easy to test whether the intruders can guarantee a score of at least one: \(Q_{LG} > 0 \Leftrightarrow \exists~q_k > 0\).

By dividing the agents into subteams according to $\bf G^*$, and by each subteam playing the two vs.\ one game against the corresponding defender pair, the intruder team guarantees the following (see Theorem~1 in \cite{shishika2020ral}):
\begin{equation}
Q \geq Q_{LG}.
\end{equation}
Note that such team strategy for the intruder and the score lower bound are not given by either MM or MIS analyses.
Note also that this score lower bound is independent of the defender strategy.

For the defender team strategy we also use the ``independent'' intruder winning region, $\awinri$, defined in \eqref{eq:independent_awinr}.
We define an extended version of the local game score:
\begin{equation}
    \hat{q}_k = q_k + \hat{n}_A^k,
\end{equation}
where $\hat{n}_A^k$ denotes the number of intruders in the corresponding paired-defense region $\dwinrp$.

The LGR defense policy developed in \cite{shishika2020ral} takes the following steps:
\begin{itemize}
    \item[1)] Remove/ignore $Q_{LG}$ uncapturable intruders from the game, so that the defenders can play a virtual game with $Q_{LG}=0$ (i.e., $q_k=0, \forall\;k$). Identification of these intruders is presented as Alg.~2 in \cite{shishika2020ral}.
    \item[2)] For each region with $\hat{q}_k\geq 1$, assign corresponding defender pair to one of the intruders in $\dwinrp$. A greedy algorithm for this two vs.\ one assignments is presented as Alg.~3 in \cite{shishika2020ral}.
    \item[3)] Perform Maximum Matching to assign one vs.\ one defense for the remaining intruders and defenders (Alg.~4 in \cite{shishika2020ral}).
\end{itemize}

We show in \cite{shishika2020ral} that the above procedure has polynomial time complexity.
In addition, we show that if $\hat{q}_k \leq 1,\forall k$ after the removal of uncapturable intruders in the first step, then the LGR defense policy guarantees the following (Theorem~3 in \cite{shishika2020ral}):
\begin{equation}
    Q \leq Q_{LG}.
\end{equation}
Together with the previous lower bound provided by the intruder team, this result proves the optimality of this defense policy in a sense that it constitutes a saddle-point equilibrium.

As long as the intruder team sticks to their equilibrium strategy, the defender team cannot reduce the score by deviating from LGR defense policy.
This optimality indirectly proves the following result:
\begin{equation}
    Q \leq Q_{LG} \leq Q_{MIS} \leq Q_{MM}.
    \label{eq:tightness}
\end{equation}
We can in fact construct a case where $Q_{LG}<Q_{MIS}$ as discussed in the simulation section.

\subsection{Discussions}
We discuss the strengths and weaknesses of the three approaches introduced in this section: MM, MIS, and LGR defense policies.

The MM assignment has the best computational efficiency, and it is also the simplest approach to use.\footnote{Note that we solely account for the multi-agent assignment aspect and not the individual winning regions here.}
The only necessary information from the agent-level game is the pair-wise win/loss information for all defender-intruder pairs.
This simplicity allows us to use the MM assignment even when the defenders have different speed limits or even different dynamics, since the analysis comes down to the individual performance.
Therefore the MM approach is also the most extensible one as well.
All of the above strengths come at the cost of suboptimal defender behavior, due to the absence of cooperative two vs.~one defense.

The MIS assignment improves the score bound at the cost of computational complexity.
It is still simple to set up and extensible since we are only augmenting the one vs.\ one results with additional two vs.\ one results and posing it as an existing combinatorial optimization problem.
Again, the biggest drawback is the computational complexity, which makes this approach suitable only for small problems.

The LGR approach has two main strengths.
First, the LGR defense policy gives the tightest score bound, as described in $\eqref{eq:tightness}$, and it actually constitutes a Nash equilibrium \cite{shishika2020ral}.
The second strength of the LGR analysis is that it also provides a lower bound on the score: $Q_{LG}$.
This is in contrast to MM or MIS approaches that only provide score upper bounds.
The intruder team strategy as well as the score lower bound given by the LGR analysis are useful tools in assessing the performance of defense systems.

Even with these strengths, the LGR algorithm is tractable in a sense that the complexity grows polynomially with the number of agents.
The numerical comparison of the score bounds $Q_{MM}$ and $Q_{LG}$ is presented in \cite{shishika2020ral}.
Also, the time complexity of all three approaches are provided in the Appendix of \cite{shishika2020ral}.
The down side of the LGR approach lies in a relatively sophisticated formulation related to the subteam definitions.
In addition, the current theory only accommodates defender teams with homogeneous speed limits, and the extension to higher-order dynamics will be non-trivial.

In summary, the MM defense strategy should be considered when simplicity and extensibility are important.
The LGR defense strategy should be used when optimality is important.
Finally, if the intruder team strategy and/or score lower bound are useful, the LGR analysis provides these information.
}
\fi

\vspace{-5 pt}
\section{Simulation Examples} \label{sec:simulation}
This section demonstrates the theoretical results through numerical examples.
\rev{All the examples use the perimeter shape parameterized as follows: $[x,y] = [a\cos\theta,b\sin\theta]$, where $[a,b]=[5,2],[2,2],[2,3]$ and $[5,3]$ for the polar angles $\theta\in[0,\frac{\pi}{2}]$, $\theta\in[\frac{\pi}{2},\pi]$, $\theta\in[\pi,\frac{3\pi}{2}]$, and $\theta\in[\frac{3\pi}{2},2\pi]$.
Note however that such parameterization is not necessary to apply the results of this paper.
The perimeter curve maybe given as a series of sample points, or as a set of vertices of a polygon.
The number of those data points linearly affect the overall computational complexity through the search for the breach point.
}

\subsection{One vs.\ One Game}
We verify the results in Sec.~\ref{sec:twoplayergame} by testing both optimal and suboptimal intrusion strategies.
We select the speed ratio to be $\nu = 0.8$ and start the game in the intruder-winning configuration.
\begin{figure}[t]
\centering
\includegraphics[width=.48\textwidth]
{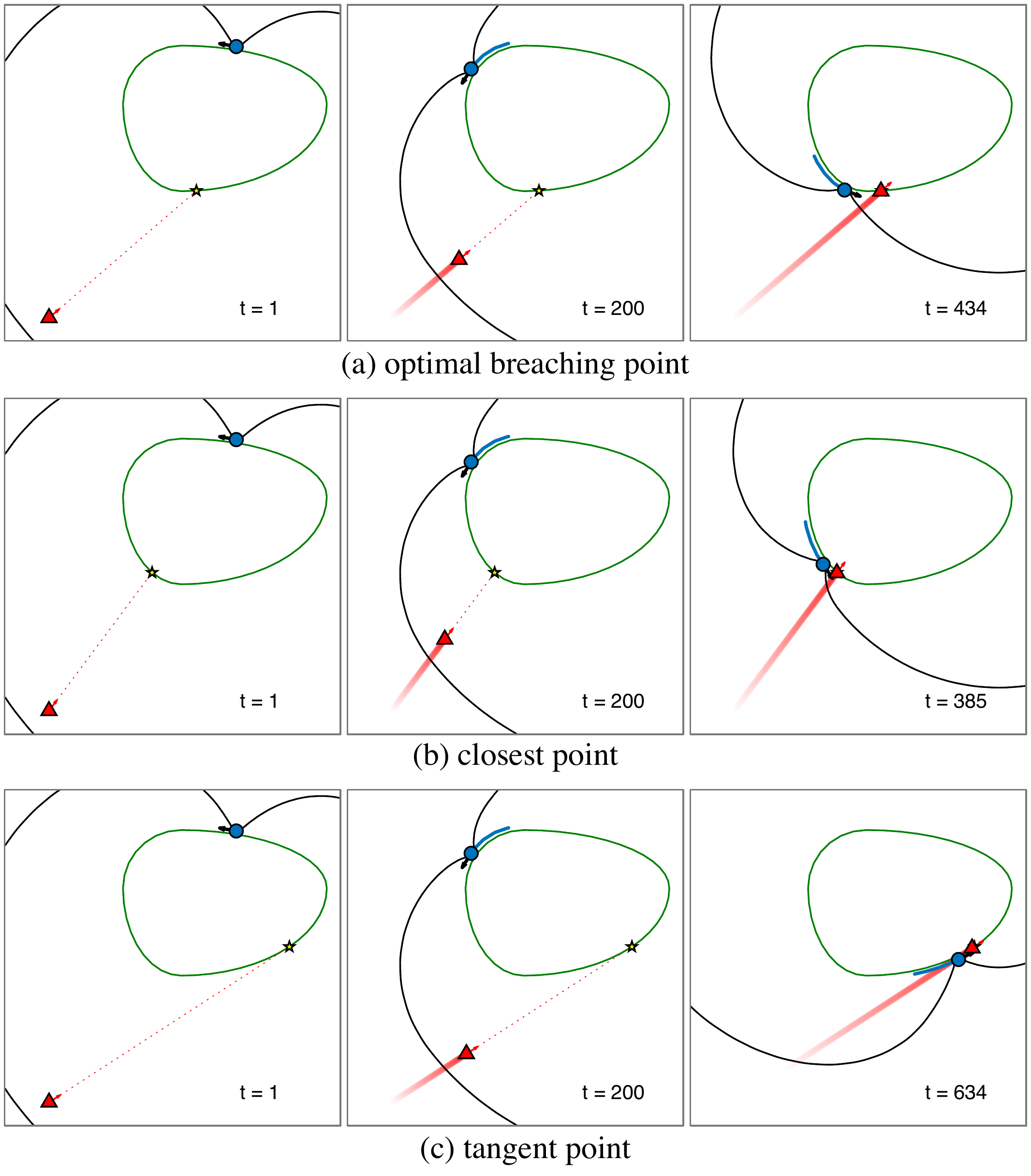}
\caption{
Simulation snapshots of one vs.\ one game with $\nu=0.8$.
(a) Intruder behavior using the correct speed ratio.
(b) Intruder behavior using $\nu=0.01$.
(c) Intruder behavior using $\nu=1$.
}
\label{fig:sim_example_1v1}
\end{figure}
Fig.~\ref{fig:sim_example_1v1}a shows the simulation snapshots when the intruder takes the optimal strategy, whereas Fig.~\ref{fig:sim_example_1v1}b and c show the cases when the intruder behaves suboptimally.
\revrev{The computation time of the strategies was 0.2 ms for an implementation in Matlab running on a laptop with a Core i7-7820HQ processor with 16 GB of memory.}

By inspecting the right most column, we can compare the performance in terms of two metrics.
First, the distance between the defender and the intruder at this time is the \emph{safe distance} considered in Sec.~\ref{sec:optimality1v1}.
We can see that the intruder achieves the largest safe distance with the optimal strategy in Fig.~\ref{fig:sim_example_1v1}a.

Next, notice the difference in the time the intruder reaches the perimeter.
By sacrificing the safe distance, the closest-point strategy in Fig.~\ref{fig:sim_example_1v1}b shows an improved performance in terms of the arrival time.
This strategy also has the property of being \emph{open-loop} type, since the closest point on the perimeter is completely independent of the defender's position or its behavior.
However, note that this strategy does not always guarantee intruder's win even if the game starts in the intruder winning configuration.
Specifically, when the intruder starts on the barrier, only the optimal strategy guarantees its win.

The tangent-point strategy in Fig.~\ref{fig:sim_example_1v1}c shows the opposite effect in the time of arrival.
By sacrificing the safe distance, this strategy delays the time the game ends, which may become relevant in a multi-player game where it tries to keep the defender away from other intruders.
The result of this example highlights the fact that the optimal strategy will be different if the intruder's objective is to delay the capture as much as possible.

We omit the demonstration of the suboptimal defender strategy since it is already shown clearly with Fig.~\ref{fig:intruder_winning} in Sec.~\ref{sec:winning_regions}.

\subsection{Multiplayer Game}
\if\ARXIVversion1
\begin{figure*}[t]
\centering
\includegraphics[width=.98\textwidth]
{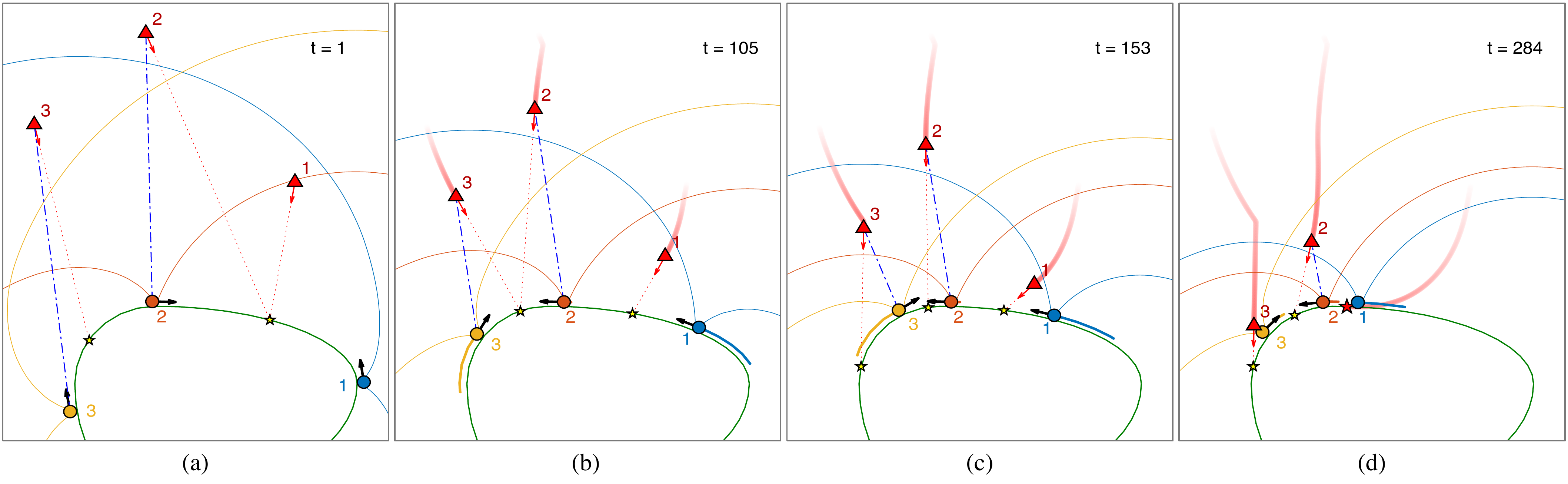}
\caption{
Simulation snapshots of MM defense.
}
\label{fig:sim_example_mm}
\end{figure*}
\fi
\begin{figure*}[t]
\centering
\includegraphics[width=.98\textwidth]
{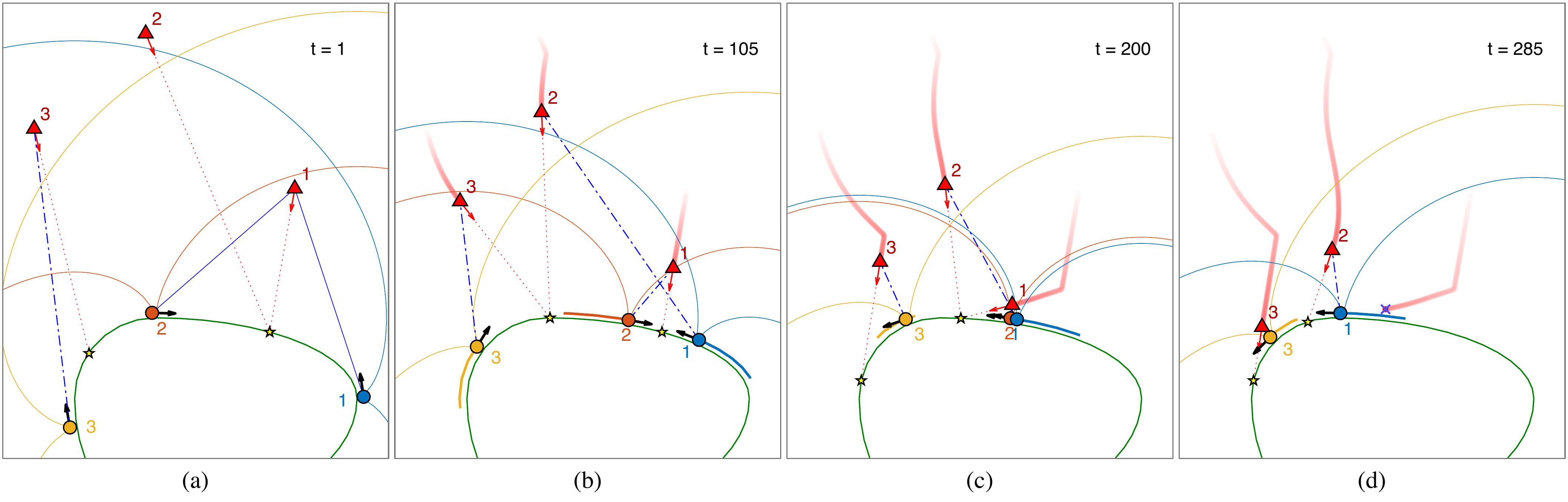}
\caption{
Simulation snapshots of MIS defense.
}
\label{fig:sim_example_mis}
\end{figure*}
\if\ARXIVversion0
The case with multiple defenders and intruders are shown in Fig.~\ref{fig:sim_example_mis}.\footnote{also see \texttt{https://youtu.be/h0{\char`_}VqJbNsQc} for the animated version.}
The small yellow stars indicate each intruder's breaching point, the dash-dotted lines indicate the one vs.\ one assignments, and the solid blue lines indicate the two vs.\ one assignments.

The intruders are performing independently greedy behavior: i.e., there is no team coordination\footnote{see \cite{shishika2020ral} for a coordinated team strategy of the intruders.}.
Each intruder finds the closest pair of defenders that contains itself in the ``relevant region'', defined by the area between the two afferent surfaces (see Fig.~\ref{fig:regions2v1}a). 
Then the intruder plays the two vs.\ one game against the pair.
For example, in Fig.~\ref{fig:sim_example_mis}a both $A_1$ and $A_2$ are located in the relevant region against the pair ($D_1,D_2$), and therefore move towards the mid point between the two defenders.
The kink in the path of $A_3$ (see Fig.~\ref{fig:sim_example_mis}d), is generated due to the switching from the midpoint between ($D_2,D_3$) to the one between ($D_3,D_1$).

In this example, the defender team is employing Maximum Independent Set (MIS) strategy \cite{shishika2018cdc}.
The pair $(D_1,D_2)$ initially plays the two vs.\ one game against $A_1$.
However, at time $t=105$, the intruder $A_1$ moves into $\dwinr(D_2)$, which frees $D_1$ from the two vs.\ one game and allows it to perform a one vs.\ one game against $A_2$.
At this point, the score upperbound is $\Qmis=0$, and the defender team guarantees that no intruder scores.
The full discussion and comparison of different team strategies are presented in the unabridged version \cite{shape_long}.
\fi

\if\ARXIVversion1
The example provided in Sec.~\ref{sec:team_strategy} (Fig.~\ref{fig:MIS}a) showed a trivial case in which the MIS defense outperforms MM defense, i.e., a case where $\Qmis < \Qmm$.
Here, we show an example where the two strategies initially have the same guarantee $\Qmm=\Qmis$, but only MIS actually performs better than the initially provided bound.

Simulation snapshots of a three vs.\ three scenario are shown in Fig.~\ref{fig:sim_example_mm} and \ref{fig:sim_example_mis} for MM defense and MIS defense respectively.\footnote{Also see \texttt{https://youtu.be/h0{\char`_}VqJbNsQc} for the animated version.}
The small yellow stars indicate each intruder's breaching point, the dash-dotted lines indicate the one vs.\ one assignments, and the solid blue lines indicate the two vs.\ one assignments.

The intruders are performing independently greedy behavior: i.e., there is no team coordination.\footnote{See \cite{shishika2018cdc} for a coordinated team strategy of the intruders.}
Each intruder finds the closest pair of defenders that contains itself in the ``relevant region'', defined by the area between the two afferent surfaces (see Fig.~\ref{fig:regions2v1}a). 
Then the intruder plays the two vs.\ one game against the pair.
For example, in Fig.~\ref{fig:sim_example_mm}a both $A_1$ and $A_2$ are located in the relevant region against the pair ($D_1,D_2$), and therefore move towards the mid point between the two defenders.

Once the intruder converges on the afferent surface of a defender, the relevant region may start switching frequently.
For example, at time $t=105$, the intruder $A_2$ is already on the afferent surface of defender $D_2$.
Depending on the side of a small deviation, the relevant pair for $A_2$ switches between ($D_1,D_2$) and ($D_2,D_3$).
Such switching causes the intruder to follow a zigzag path towards defender $D_2$.
To avoid such degenerate behavior, we add a small bias towards ccw direction when the intruder selects the pair, which is why $A_2$ selects the pair ($D_2,D_3$).
The kink in the path of $A_3$ (see Fig.~\ref{fig:sim_example_mm}d), is generated due to the switching from the midpoint between ($D_2,D_3$) to the one between ($D_3,D_1$).

The MM assignment shown in Fig.~\ref{fig:sim_example_mm}, has two valid edges giving $\numcapture = 2$ and $\Qmm=N_A-\numcapture = 1$.
Since this MM assignment does not specify any behavior to the unassigned defenders, we also consider a secondary matching between the unassigned intruders and defenders.
Defender $D_1$ gets this secondary assignment towards $A_1$, which is why $D_1$ moves ccw.
As the $\Qmm$ from the MM analysis expected, $A_1$ scores a point (Fig.~\ref{fig:sim_example_mm}d).

The MIS assignment shown in Fig.~\ref{fig:sim_example_mis} also has $\numcapture=2$ at the beginning, only guaranteeing $Q\leq \Qmis=1$.
\rev{For this small problem, the computation time of the MIS defense strategy was 5 ms.}
The pair $(D_1,D_2)$ initially plays the two vs.\ one game against $A_1$.
However, at time $t=105$, the intruder $A_1$ moves into $\dwinr(D_2)$, which frees $D_1$ from the two vs.\ one game and allows it to perform a one vs.\ one game against $A_2$.
At this point, the score upperbound has changed to $\Qmis=0$, and the defender team guarantees that no intruder scores.

Although the score bound provided by $\Qmis$ is tighter than $\Qmm$ (see Sec.~\ref{sec:team_strategy}), this example highlights that it may still not be the smallest upper bound.
Specifically, the MIS analysis could not predict the outcome $Q=0$ from the initial configuration.
We also note that the MIS assignment is non-unique; in fact, it could have selected the same edge set as the MM assignment in this example, because they both have the same cardinality.
In other words, the two assignments are equally good in the instantaneous analysis.
However, only the assignment shown in Fig.~\ref{fig:sim_example_mis} leads to the capture of all intruders.

\rev{
If we use the LGR defense strategy, we have $q_k=0$ for all the regions, and thus we have $Q_{LG}=0$.  This implies that all intruders will be captured.  The assignment will be the same as the one in Fig.~\ref{fig:sim_example_mis}.  However, what is important is that the LGR strategy always makes this ``correct'' decision.  This is one of the reasons behind the performance gap between MIS and LGR defense strategies. In addition, unlike the MIS defense strategy, LGR analysis could predict $Q=0$ from the initial configuration, showing that it is a more accurate estimate of the game outcome.
}
\fi

\vspace{-5 pt}
\section{Conclusion}\label{sec:conclusion}
\vspace{-5 pt}
We study a variant of the reach-avoid game with the defenders constrained to move on the perimeter of the target region.
The intruders try to score by breaching the perimeter while the defender team tries to minimize the score by intercepting them.
The one vs.\ one game is solved analytically for arbitrary convex shapes, which provides the intruder's optimal breaching point and the defender's optimal direction of motion.
The derived strategies are at an equilibrium in terms of the safe distance (in the attacker-winning scenario) and the largest margin (in the defender-winning scenario).
The two vs.\ one game is also solved analytically, and it highlights the benefit of cooperation among the defenders.
Specifically, two defenders can team up to perform a pincer maneuver to reduce the intruder-winning region.
\if \ARXIVversion1
\rev{
Finally, we introduce and discuss various team defense strategies that leverage the results from one vs.\ one and two vs.\ one games.
}
\fi
\if \ARXIVversion0
These results can be directly used in the team strategies for the multi-player games as described in the unabridged version \cite{shape_long}.
\fi

%
\vspace{-5 pt}
\begin{ack}                               
\vspace{-5 pt}
We gratefully acknowledge useful discussions with Chris Kroninger, Ken Hayashima, and Alexander Von Moll.
\end{ack}

\vspace{-5 pt}
\bibliographystyle{plain}        
\bibliography{Automatica_shape}           

\begin{thebibliography}{10}

\bibitem{Agharkar2014}
Pushkarini Agharkar and Francesco Bullo.
\newblock {Vehicle routing algorithms to intercept escaping targets}.
\newblock {\em Proc. Amer. Control Conf. (ACC)}, pages 952--957, 2014.

\bibitem{Bakolas2012}
Efstathios Bakolas and Panagiotis Tsiotras.
\newblock {Relay pursuit of a maneuvering target using dynamic Voronoi
  diagrams}.
\newblock {\em Automatica}, 48(9):2213--2220, 2012.

\bibitem{TamerBasar}
Tamer Basar and Geert~Jan Olsder.
\newblock {\em {Dynamic Noncooperative Game Theory, 2nd Edition}}.
\newblock Society for Industrial and Applied Mathematics, 2011.

\bibitem{Bopardikar2009}
Shaunak~D. Bopardikar, Francesco Bullo, and Jo{\~{a}}o~P. Hespanha.
\newblock {A cooperative homicidal chauffeur game}.
\newblock {\em Automatica}, 45(7):1771--1777, 2009.

\bibitem{Chen2014b}
Mo~Chen, Zhengyuan Zhou, and Claire~J. Tomlin.
\newblock {A path defense approach to the multiplayer reach-avoid game}.
\newblock {\em IEEE Conf. Decis. Control (CDC)}, pages 2420--2426, 2014.

\bibitem{Chen2014a}
Mo~Chen, Zhengyuan Zhou, and Claire~J. Tomlin.
\newblock {Multiplayer reach-avoid games via low dimensional solutions and
  maximum matching}.
\newblock {\em Proc. Amer. Control Conf. (ACC)}, pages 1444--1449, 2014.

\bibitem{chen2017}
Mo~Chen, Zhengyuan Zhou, and Claire~J. Tomlin.
\newblock {Multiplayer reach-avoid games via pairwise outcomes}.
\newblock {\em IEEE Trans. Autom. Control}, 62(3):1451--1457, mar 2017.

\bibitem{Chung2011}
Timothy~H Chung and Geoffrey~A Hollinger.
\newblock {Search and pursuit-evasion in mobile robotics}.
\newblock {\em Auton. Robot.}, 31(4):299--316, 2011.

\bibitem{corliss1977root}
George Corliss.
\newblock Which root does the bisection algorithm find?
\newblock {\em Siam Review}, 19(2):325--327, 1977.

\bibitem{filippov2013}
Aleksei~Fedorovich Filippov.
\newblock {\em Differential equations with discontinuous righthand sides:
  control systems}, volume~18.
\newblock Springer Science \& Business Media, 2013.

\bibitem{Fisac2015a}
Jaime~F. Fisac, Mo~Chen, Claire~J. Tomlin, and S.~Shankar Sastry.
\newblock {Reach-avoid problems with time-varying dynamics, targets and
  constraints}.
\newblock {\em Proc. 18th Int. Conf. Hybrid Sys. Comp. Control (ACM)}, pages
  11--20, 2015.

\bibitem{Fisac2015}
Jaime~F. Fisac and S.~Shankar Sastry.
\newblock {The pursuit-evasion-defense differential game in dynamic constrained
  environments}.
\newblock {\em IEEE Conf. Decis. Control (CDC)}, pages 4549--4556, 2015.

\bibitem{Foley1974a}
M~Foley and W~Schmitendorf.
\newblock {A class of differential games with two pursuers versus one evader}.
\newblock {\em IEEE Trans. Autom. Control}, 19(3):239--243, 1974.

\bibitem{Fuchs2010}
Zachariah~E. Fuchs, Pramod~P. Khargonekar, and Johnny Evers.
\newblock {Cooperative defense within a single-pursuer, two-evader pursuit
  evasion differential game}.
\newblock {\em IEEE Conf. Decis. Control (CDC)}, pages 3091--3097, 2010.

\bibitem{Garcia2015}
Eloy Garcia, David~W. Casbeer, Khanh Pham, and Meir Pachter.
\newblock {Cooperative aircraft defense from an attacking missile}.
\newblock {\em J. Guid. Control Dyn}, 38(8):1510--1520, 2015.

\bibitem{garcia2019cooperative}
Eloy Garcia, David~W Casbeer, Alexander Von~Moll, and Meir Pachter.
\newblock Cooperative two-pursuer one-evader blocking differential game.
\newblock In {\em Proc. Amer. Control Conf. (ACC)}, pages 2702--2709. IEEE,
  2019.

\bibitem{garcia2020multiple}
Eloy Garcia, David~W Casbeer, Alexander Von~Moll, and Meir Pachter.
\newblock Multiple pursuer multiple evader differential games.
\newblock {\em IEEE Trans. Autom. Control}, 2020.

\bibitem{Garcia2017}
Eloy Garcia, Zachariah~E. Fuchs, Dejan Milutinovic, David~W. Casbeer, and Meir
  Pachter.
\newblock {A Geometric Approach for the Cooperative Two-Pursuer One-Evader
  Differential Game}.
\newblock {\em IFAC-PapersOnLine}, 50(1):15209--15214, 2017.

\bibitem{garcia2019coastline}
Eloy Garcia, Alexander Von~Moll, David~W Casbeer, and Meir Pachter.
\newblock Strategies for defending a coastline against multiple attackers.
\newblock In {\em IEEE Conf. Decis. Control (CDC)}, pages 7319--7324, 2019.

\bibitem{Huang2011icra}
Haomiao Huang, Jerry Ding, Wei Zhang, and Claire~J. Tomlin.
\newblock {A differential game approach to planning in adversarial scenarios: A
  case study on capture-the-flag}.
\newblock {\em IEEE Int. Conf. Rob. Autom. (ICRA)}, pages 1451--1456, 2011.

\bibitem{Huang2011a}
Haomiao Huang, Wei Zhang, Jerry Ding, Du{\v{s}}an~M. Stipanovi{\'{c}}, and
  Claire~J. Tomlin.
\newblock {Guaranteed decentralized pursuit-evasion in the plane with multiple
  pursuers}.
\newblock {\em IEEE Conf. Decis. Control (CDC)}, pages 4835--4840, 2011.

\bibitem{Isaacs}
Rufus Isaacs.
\newblock {\em {Differential games: A mathematical theory with applications to
  warfare and pursuit, control and optimization}}.
\newblock Courier Corporation, 1999.

\bibitem{Kerns2014}
Andrew~J. Kerns, Daniel~P. Shepard, Jahshan~A. Bhatti, and Todd~E. Humphreys.
\newblock {Unmanned aircraft capture and control via GPS spoofing}.
\newblock {\em J. Field Rob.}, 31(4):617--636, 2014.

\bibitem{Kim2007a}
Tae~Hyoung Kim and Toshiharu Sugie.
\newblock {Cooperative control for target-capturing task based on a cyclic
  pursuit strategy}.
\newblock {\em Automatica}, 43(8):1426--1431, 2007.

\bibitem{KleinbergAlgorithm}
Jon Kleinberg and Eva Tardos.
\newblock {\em {Algorithm design}}.
\newblock Addison Wesley, 2006.

\bibitem{Liang2019}
Li~Liang, Fang Deng, Zhihong Peng, Xinxing Li, and Wenzhong Zha.
\newblock {A differential game for cooperative target defense}.
\newblock {\em Automatica}, 102:58--71, 2019.

\bibitem{liu2014evasion}
Shih-Yuan Liu, Zhengyuan Zhou, Claire Tomlin, and J~Karl Hedrick.
\newblock Evasion of a team of dubins vehicles from a hidden pursuer.
\newblock In {\em IEEE Int. Conf. Rob. Autom. (ICRA)}, pages 6771--6776, 2014.

\bibitem{liu2013evasion}
Shih-Yuan Liu, Zhengyuan Zhou, Claire Tomlin, and Karl Hedrick.
\newblock Evasion as a team against a faster pursuer.
\newblock In {\em IEEE Proc. Amer. Control Conf. (ACC)}, pages 5368--5373,
  2013.

\bibitem{Makkapati2018}
Venkata~Ramana Makkapati, Wei Sun, and Panagiotis Tsiotras.
\newblock {Optimal Evading Strategies for Two-Pursuer/One-Evader Problems}.
\newblock {\em J. Guid. Control Dyn}, 41(4):851--862, 2018.

\bibitem{Makkapati2019}
Venkata~Ramana Makkapati and Panagiotis Tsiotras.
\newblock {Optimal Evading Strategies and Task Allocation in Multi-player
  Pursuit–Evasion Problems}.
\newblock {\em Dynamic Games and Applications}, pages 1--20, 2019.

\bibitem{Mitchell2014}
Robert Mitchell and Ing~Ray Chen.
\newblock {Adaptive intrusion detection of malicious unmanned air vehicles
  using behavior rule specifications}.
\newblock {\em IEEE Trans. Syst. Man Cybern.: Syst.}, 44(5):593--604, 2014.

\bibitem{Oyler2016}
Dave~W. Oyler, Pierre~T. Kabamba, and Anouck~R. Girard.
\newblock {Pursuit-evasion games in the presence of obstacles}.
\newblock {\em Automatica}, 65:1--11, 2016.

\bibitem{pachter2019singular}
Meir Pachter, Alexander Von~Moll, Eloy Garcia, David~W Casbeer, and Dejan
  Milutinovi{\'c}.
\newblock Singular trajectories in the two pursuer one evader differential
  game.
\newblock In {\em 2019 International Conference on Unmanned Aircraft Systems
  (ICUAS)}, pages 1153--1160. IEEE, 2019.

\bibitem{Pasqualetti2012}
Fabio Pasqualetti, Antonio Franchi, and Francesco Bullo.
\newblock {On cooperative patrolling: Optimal trajectories, complexity
  analysis, and approximation algorithms}.
\newblock {\em IEEE Trans. Rob.}, 28(3):592--606, 2012.

\bibitem{Pierson2017}
Alyssa Pierson, Zijian Wang, and Mac Schwager.
\newblock {Intercepting rogue robots: An algorithm for capturing multiple
  evaders with multiple pursuers}.
\newblock {\em IEEE Rob. Autom. Lett.}, 2(2):530--537, 2017.

\bibitem{Rubinsky2014}
Sergey Rubinsky and Shaul Gutman.
\newblock {Three-Player Pursuit and Evasion Conflict}.
\newblock {\em J. Guid. Control Dyn}, 37(1):98--110, 2014.

\bibitem{Scott2018}
William~L. Scott and Naomi~E. Leonard.
\newblock {Optimal evasive strategies for multiple interacting agents with
  motion constraints}.
\newblock {\em Automatica}, 94:26--34, 2018.

\bibitem{Selvakumar2019}
Jhanani Selvakumar and Efstathios Bakolas.
\newblock {Feedback strategies for a reach-avoid game with a single evader and
  multiple pursuers}.
\newblock {\em IEEE Trans. Cybern.}, PP:1--12, 2019.

\bibitem{shishika2018cdc}
Daigo Shishika and Vijay Kumar.
\newblock {Local-game decomposition for multiplayer perimeter-defense problem}.
\newblock In {\em IEEE Conf. Decis. Control (CDC)}, pages 2093--2100, 2018.

\bibitem{shishika2019team}
Daigo Shishika, James Paulos, Michael~R Dorothy, M~Ani Hsieh, and Vijay Kumar.
\newblock Team composition for perimeter defense with patrollers and defenders.
\newblock In {\em IEEE Conf. Decis. Control (CDC)}, pages 7325--7332, 2019.

\bibitem{shishika2020ral}
Daigo Shishika, James Paulos, and Vijay Kumar.
\newblock Cooperative team strategies for multi-player perimeter-defense games.
\newblock {\em IEEE Rob. Autom. Lett.}, 5(2):2738--2745, 2020.

\bibitem{takei2014efficient}
Ryo Takei, Richard Tsai, Zhengyuan Zhou, and Yanina Landa.
\newblock An efficient algorithm for a visibility-based surveillance-evasion
  game.
\newblock {\em Comm. in Math. Sci.}, 12(7):1303--1327, 2014.

\bibitem{von2019multi}
Alexander Von~Moll, David Casbeer, Eloy Garcia, Dejan Milutinovi{\'c}, and Meir
  Pachter.
\newblock The multi-pursuer single-evader game.
\newblock {\em J. Intel. Rob. Syst.}, 96(2):193--207, 2019.

\bibitem{von2019multiple}
Alexander Von~Moll, Eloy Garcia, David Casbeer, M~Suresh, and Sufal~Chandra
  Swar.
\newblock Multiple-pursuer, single-evader border defense differential game.
\newblock {\em J. Aero. Info. Syst.}, pages 1--10, 2019.

\bibitem{von2020robust}
Alexander Von~Moll, Meir Pachter, Eloy Garcia, David Casbeer, and Dejan
  Milutinovi{\'c}.
\newblock Robust policies for a multiple-pursuer single-evader differential
  game.
\newblock {\em Dynamic Games and Applications}, 10(1):202--221, 2020.

\bibitem{yan2019three}
Rui Yan, Xiaoming Duan, Zongying Shi, Yisheng Zhong, and Francesco Bullo.
\newblock Maximum-matching capture strategies for 3d heterogeneous multiplayer
  reach-avoid games.
\newblock {\em arXiv preprint arXiv:1909.11881}, 2019.

\bibitem{yan2017escape}
Rui Yan, Zongying Shi, and Yisheng Zhong.
\newblock Escape-avoid games with multiple defenders along a fixed circular
  orbit.
\newblock In {\em 13th IEEE Int. Conf. Control \& Autom. (ICCA)}, pages
  958--963. IEEE, 2017.

\bibitem{yan2018twovsone}
Rui Yan, Zongying Shi, and Yisheng Zhong.
\newblock Reach-avoid games with two defenders and one attacker: An analytical
  approach.
\newblock {\em IEEE Trans. Cybern.}, 49(3):1035--1046, 2018.

\bibitem{yan2019analytical}
Rui Yan, Zongying Shi, and Yisheng Zhong.
\newblock Task assignment for multiplayer reach--avoid games in convex domains
  via analytical barriers.
\newblock {\em IEEE Trans. Rob.}, 36(1):107--124, 2019.

\bibitem{zhou2018efficient}
Zhengyuan Zhou, Jerry Ding, Haomiao Huang, Ryo Takei, and Claire Tomlin.
\newblock Efficient path planning algorithms in reach-avoid problems.
\newblock {\em Automatica}, 89:28--36, 2018.

\bibitem{Zhou2012general}
Zhengyuan Zhou, Ryo Takei, Haomiao Huang, and Claire~J Tomlin.
\newblock A general, open-loop formulation for reach-avoid games.
\newblock In {\em IEEE Conf. Decis. Control (CDC)}, pages 6501--6506, 2012.

\bibitem{Zhou2016}
Zhengyuan Zhou, Wei Zhang, Jerry Ding, Haomiao Huang, Du{\v{s}}an~M.
  Stipanovi{\'{c}}, and Claire~J. Tomlin.
\newblock {Cooperative pursuit with Voronoi partitions}.
\newblock {\em Automatica}, 72:64--72, 2016.

\end{thebibliography}




\end{document}